\documentclass[twocolumn,english]{article}

\usepackage[T1]{fontenc}
\usepackage[utf8]{inputenc}
\usepackage[a4paper]{geometry}
\usepackage{color}
\usepackage{babel}
\usepackage{bm}
\usepackage{amsmath}
\usepackage{amssymb}
\usepackage{graphicx}
\usepackage[unicode=true]
 {hyperref}

\makeatletter
\usepackage{babel}
\usepackage{xcolor}
\definecolor{myred}{rgb}{0.66, 0.15, 0.15}
\definecolor{darkgreen}{rgb}{0.0, 0.5, 0.0}

\usepackage{hyperref}          
\hypersetup{
     colorlinks   = true,
     citecolor    = myred,
     urlcolor     = myred,
     urlbordercolor = myred,
     linkcolor = myred,
}

\usepackage{flushend}
\usepackage{cuted}
\usepackage{cite}

\newcommand{\seeAppendix}[1]{(see Appendix #1)}
\newcommand{\supplementalBibliography}{}

\newcommand{\uzero}{f_{0}}        
\newcommand{\uinfty}{f_{\infty}}  

\newcommand{\QDTetabar}{\bar{\eta}}
\newcommand{\QDTetabarkla}{\bar{\eta}} 

\newcommand{\QDTX}{B_{\text{o}}}       
\newcommand{\QDTXkla}{B_{a}} 

\newcommand{\etak}{\eta} 
\newcommand{\etaok}{\eta_{\text{o}}} 
\newcommand{\etabgk}{\eta_{\text{bg}}} 

\newcommand{\QDTuok}{\bar{u}_{\text{o}}}       
\newcommand{\QDTvok}{\bar{v}_{\text{o}}}       
\newcommand{\QDTvokplus}{\bar{v}_{\text{o}}^{+}}       

\newcommand{\QDTuk}{\bar{u}_{a}}       
\newcommand{\QDTvk}{\bar{v}_{a}}       
\newcommand{\QDTvkplus}{\bar{v}_{a}^{+}}       
\newcommand{\QDTukl}{\QDTukl^{k,\ell}}       

\newcommand{\QDTukZero}{\bar{u}_{0}}       
\newcommand{\QDTukInfi}{\bar{u}_{\infty}}     

\newcommand{\QDTa}{a}
\newcommand{\QDTetakla}{\eta_{a}} 
\newcommand{\QDTDkla}{D_{a}} 
\newcommand{\QDTPkla}{P_{a}} 
\newcommand{\QDTbkla}{b_{a}} 
\newcommand{\QDTPklaplus}{P_{a}^{+}} 
\newcommand{\QDTbklaplus}{b_{a}^{+}} 
\newcommand{\QDTbklplus}{b^{+}} 

\newcommand{\QDTDklo}{D_{\text{o}}} 
\newcommand{\QDTPklo}{P_{\text{o}}} 
\newcommand{\QDTbklo}{b_{\text{o}}} 

\newcommand{\QDTDklzero}{D_{0}} 
\newcommand{\QDTZ}{A}
\newcommand{\QDTZk}{\QDTZ} 
\newcommand{\QDTetazero}{\eta_{0}} 
\newcommand{\QDTetainf}{\eta_{\infty}} 

\newcommand{\QDTalpha}{\gamma}
\newcommand{\QDTlambdakappa}{\lambda(\kappa)}

\newcommand{\QDTgklo}{g_{\text{o},\ell}} 
\newcommand{\QDTgkloplus}{g^{+}_{\text{o},\ell}} 

\newcommand{\phio}{\phi_{\text{o}}}
\newcommand{\phic}{\phi_{\text{c}}}
\newcommand{\phib}{\phi_{\text{b}}}
\newcommand{\phiobar}{\bar{\phi}_{\text{o}}^{E,\hat{k}}}
\newcommand{\phiobarp}{\bar{\phi}_{\text{o}}^{E^\prime,\hat{k^\prime}}}
\newcommand{\phibg}{\phi_{\text{bg}}}
\newcommand{\phiol}{\phi_{\text{o},\ell}} 
\newcommand{\phiobarl}{\bar{\phi}_{\text{o},\ell}} 
\newcommand{\uol}{u_{\text{o}}} 
\newcommand{\ucl}{u_{\text{c}}}
\newcommand{\ubg}{u_{\text{bg}}}

\newcommand{\myX}{X}
\newcommand{\myY}{Y}
\newcommand{\myZ}{Z}

\makeatletter
\newcommand\semiHuge{\@setfontsize\semiHuge{32.72}{37.38}}
\renewcommand\section{\@startsection{section}{1}{\z@}%
                                 {-3.25ex\@plus -1ex \@minus -.2ex}%
                                 {1.5ex \@plus .2ex}%
                                 {\raggedright \normalfont\Large\bfseries}}
\renewcommand\subsection{\@startsection{subsection}{2}{\z@}%
                                 {-3.25ex\@plus -1ex \@minus -.2ex}%
                                 {1.5ex \@plus .2ex}%
                                 {\raggedright \normalfont\large\bfseries}}
\makeatother

\makeatother

\begin{document}
\begin{strip}\vspace{-1cm}

\begin{flushleft}
\textsf{\textbf{\huge{}Closed-channel parameters of Feshbach resonances}}{\huge\par}
\par\end{flushleft}

\textsf{\vspace{-0.5cm}
}

{\Large{}Pascal Naidon}\textsf{\textbf{\huge{}\vspace{0.2cm}
}}{\huge\par}

{\small{}Few-Body Systems Physics Laboratory, \href{https://ror.org/05tqx4s13}{RIKEN Nishina Centre},
RIKEN, Wak{ō}, 351-0198 Japan.}\textit{ }{\Large{}\vspace{-0.4cm}
}{\Large\par}

\textit{\href{mailto:pascal@riken.jp}{pascal@riken.jp}}\textsf{\textbf{\huge{}\vspace{0.2cm}
}}{\huge\par}

\today{\Large{}\vspace{0.5cm}
}{\Large\par}

This work investigates how the closed channel of a Feshbach resonance
is characterised by experimental observables. Surprisingly, it is
found that the two-body observables associated with the Feshbach resonance
can be insensitive to the properties of the closed channel. In particular,
it is impossible in this situation to determine the energy of the
bound state causing the resonance from the usual experimental data.
This is the case for all magnetic Feshbach resonances in ultracold
atoms, due to their deep two-body interaction potentials. This insensitivity
highlights a major difference with Feshbach resonances that involve
shallow interaction potentials, such as hadron resonances. It appears
however that short-range two-body correlations and three-body observables
are affected by a parameter of the closed channel called the ``closed-channel
scattering length''. A photoassociation experiment is proposed to
measure this parameter in ultracold atom systems.

\vspace{0.5cm}

\end{strip}

\section{Introduction\label{sec:Introduction}}

Feshbach resonances~\cite{Feshbach1958,Fano1961} are found in many
quantum systems, occurring whenever a continuum of states couples
to a bound state. They are particularly important in the field of
ultracold atoms, where the ``magnetic Feshbach resonances''~\cite{Chin2010,Kokkelmans2014}
have provided the possibility to control interatomic interactions
through the application of a magnetic field~\cite{Tiesinga1993,Inouye1998}.
The concept of Feshbach resonance is also used in hadron physics to
account for exotic bound states or resonances close to hadron thresholds~\cite{Hyodo2013,Dong2021,Braaten2005},
and its relevance to condensed matter systems has recently been pointed
out~\cite{Kuhlenkamp2022,Tajima2024}.

Many of the previous theoretical studies of Feshbach resonances have
been concerned with building up models that reproduce experimental
data~\cite{Chin2004,Raoult2004,Schunck2005,Gao2005,Hanna2009,Jachymski2013,Chilcott2021}.
In the present work, an opposite approach is taken by considering
which parts of the model are constrained by the observables. For this
purpose, the two-channel model describing Feshbach resonances is introduced
in Sec.~\ref{sec:Two-channel-model-1}, followed in Sec.~\ref{sec:Generic-example}
by a generic example showing an explicit dependence of observables
on the properties of the bound state responsible for the resonance.
Then, the regime where this dependence disappears is presented in
Sec.~\ref{sec:Quantum-defect-regime}, followed by a discussion of
some remarkable aspects of this regime. Finally, the possibility to
probe the closed-channel properties from short-range physics is examined
in Sec.~\ref{sec:Short-distance-physics}. 
\begin{figure*}[!t]

\hfill{}\includegraphics[width=16cm]{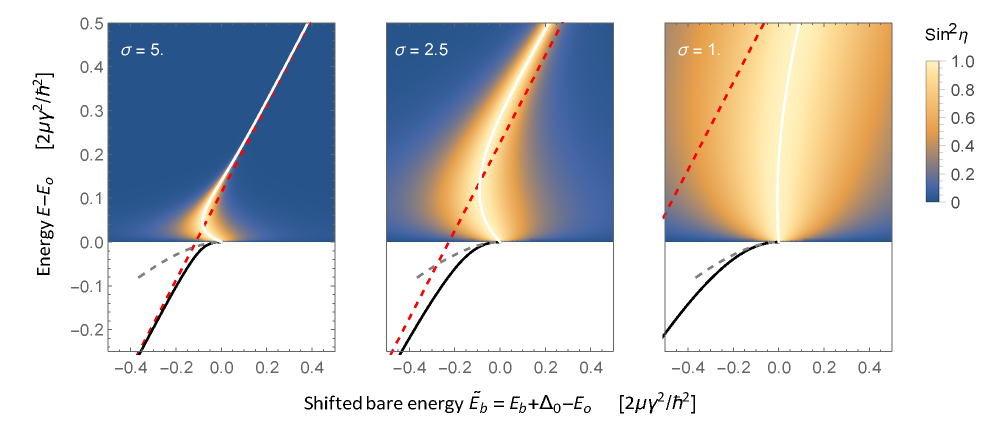}\hfill{}

\caption{\label{fig:Gaussian-model}Energy spectrum and scattering phase shift
of the non-relativistic Gaussian model of Feshbach resonance as a
function of the energy difference $\tilde{E}_{\text{b}}$ between
the shifted bare energy $E_{\text{b}}+\Delta_{0}$ and the threshold
energy $E_{\text{o}}$, for a fixed reduced width $\QDTalpha$. The
results for three different values of the closed-channel parameter
$\sigma$ (in units of $\hbar^{2}/2\mu\QDTalpha$) are shown in the
left ($\sigma=5$), middle ($\sigma=2.5$), and right ($\sigma=1$)
panels. In each panel, the shading above the continuum threshold shows
the quantity $\sin^{2}\eta$ obtained from Eqs.~(\ref{eq:scatteringPhaseShift})
and (\ref{eq:Gaussian-model-Shift+Width}), and the solid black curve
below the continuum threshold shows the dressed bound state energy
given by Eq.~(\ref{eq:Gaussian-model-BoundState}). The resonance
($\sin^{2}\eta=1$) is shown as a white curve. The dashed grey curve
shows the dressed energy in the limit $\sigma\to0$, which coincides
with the QDT Eq.~(\ref{eq:QDT_BoundState}) in the zero-range limit
given by Eq.~(\ref{eq:zeroRangeQDT2}). The dashed red line shows
the energy of the bare bound state in the closed channel causing the
resonance. One can see that the value of $\sigma$ significantly affects
the spectrum and scattering, making it possible to determine $\sigma$,
and thus the bare bound state energy, from these observables.}
\end{figure*}

\section{Two-channel model\label{sec:Two-channel-model-1}}

The following analysis is restricted to isolated resonances of a two-particle
system, i.e. a single two-body bound state $\vert\phi_{\text{b}}\rangle$
coupled to a two-body continuum. More specifically, the bound state
$\vert\phi_{\text{b}}\rangle$, which is called the ``bare bound
state'', is assumed to occur in a ``closed channel'' described
by a Hamiltonian $H_{\text{cc}}$, such that $(H_{\text{cc}}-E_{\text{b}})\vert\phi_{\text{b}}\rangle=0$,
and this closed channel is coupled through coupling terms $H_{\text{oc}}$
and $H_{\text{co}}=H_{\text{oc}}^{\dagger}$ to an ``open channel''
described by a Hamiltonian $H_{\text{oo}}$ featuring a scattering
continuum above a certain threshold $E_{\text{o}}$. The two channels
correspond to two different internal states of the particles, such
as two hyperfine states of two atoms, or two quark configurations
of a hadron. The isolated resonance theory of this two-channel model
shows that the system at energy $E$ is described by the complex
energy shift (see Appendices \ref{sec:Two-channel-model} and \ref{sec:Two-channel-isolated-resonance})
\begin{equation}
\Delta^{+}(E)\equiv\langle\phi_{\text{b}}\vert H_{\text{co}}\vert(E+i0^{+}-H_{\text{oo}})^{-1}\vert H_{\text{oc}}\vert\phi_{\text{b}}\rangle,\label{eq:Delta+}
\end{equation}
whose real and imaginary parts $\Delta$ and $-\Gamma/2$ define respectively
the shift and width of the resonance.

For energies $E$ above the open-channel threshold $E_{\text{o}}$,
the scattering properties are strongly modified for energies  around
the energy $E_{\text{b}}$ of the bare bound state. Indeed, in a certain
partial wave set by the angular momentum of $\phi_{\text{b}}$, the
scattering phase shift,
\begin{equation}
\eta(E)=\eta_{\text{bg}}(E)-\arctan\frac{\Gamma(E)/2}{E-E_{\text{b}}-\Delta(E)}\label{eq:scatteringPhaseShift}
\end{equation}
can reach unitarity (i.e. $\sin^{2}\eta=1$) at a particular energy,
corresponding to a resonant state. Here, $\eta_{\text{bg}}$ denotes
the ``background'' scattering phase shift away from that resonance.
In the following, the s wave will be considered, although other partial
waves can be treated in the same way. In this case, in the limit of
small scattering wave number $k\equiv\sqrt{2\mu(E-E_{\text{o}})}/\hbar$
(with $\mu$ being the reduced mass of the two scattering particles),
the scattering properties are governed by the s-wave scattering length
$a\equiv-\lim_{k\to0}\eta/k$. From Eq.~(\ref{eq:scatteringPhaseShift})
one finds
\begin{equation}
a=a_{\text{bg}}-\QDTalpha/\left(E_{\text{b}}+\Delta_{0}-E_{\text{o}}\right)\label{eq:scattering-length}
\end{equation}
where $a_{\text{bg}}\equiv-\lim_{k\to0}\eta_{\text{bg}}/k$ is the
background scattering length away from resonance, $\QDTalpha\equiv\lim_{k\to0}\Gamma/2k\ge0$
will be referred to as the ``'reduced width''~\cite{Timmermans1999},
and $\Delta_{0}\equiv\lim_{k\to0}\Delta(E)$ is the zero-energy shift.
Equation~(\ref{eq:scattering-length}) shows that the scattering
length can be arbitrarily large when  the bare bound state energy
$E_{\text{b}}$ shifted by $\Delta_{0}$ approaches the threshold
$E_{\text{o}}$. This divergent behaviour of the scattering length
is the basis for its control in ultracold-atomic systems by tuning
$E_{\text{b}}$, thanks to its dependence on an applied magnetic field.

For energies $E$ below the open-channel threshold $E_{\text{o}}$,
the coupled system may feature a ``dressed bound state'' (called
``Feshbach molecule''~\cite{Koehler2006,Chin2010} in the context
of ultracold-atom physics) whose energy $E_{\text{d}}$ is shifted
from the bare energy $E_{\text{b}}$ according to the formula:
\begin{equation}
E_{\text{d}}=E_{\text{b}}+\Delta(E_{\text{d}})\label{eq:BoundStateEnergy}
\end{equation}

Whether the effect of the bare bound state appears as a resonant state
above threshold, or a dressed bound state below threshold, or both,
depends on the value of the shifted energy $E_{\text{b}}+\Delta_{0}$
with respect to the threshold $E_{\text{o}}$.   It is readily
seen from Eq.\,(\ref{eq:Delta+}) that both the resonant and dressed
bound state will in general depend on the characteristics of the bare
bound state $\vert\phi_{\text{b}}\rangle$ and the coupling $H_{\text{co}}=H_{\text{oc}}^{\dagger}$.
These characteristics thus introduce ``closed-channel parameters''~\cite{Naidon2019}
into the problem. Let us now investigate how these parameters affect
observables.

\section{Generic example\label{sec:Generic-example}}

A simple example is shown in Fig.\,\ref{fig:Gaussian-model} corresponding
to a well-known non-relativistic model~\cite{JonaLasinio2008,Schmidt2012,Pricoupenko2013}
where there is no interaction between particles in the open channel,
and the coupling factor $\langle\bm{k}\vert H_{\text{oc}}\vert\phi_{\text{b}}\rangle$
is taken to be of the isotropic Gaussian type $W_{0}\exp(-k^{2}\sigma^{2}/2)$,
where $W_{0}$ and $\sigma$  constitute here the closed-channel
parameters. In this case, the shift and width above threshold can
be calculated analytically:
\begin{equation}
\Delta(k)=\Delta_{0}+\frac{\Gamma(k)}{2}\text{Im}[\text{erf}(ik\sigma)]\quad;\quad\frac{\Gamma(k)}{2}=\gamma ke^{-k^{2}\sigma^{2}}\label{eq:Gaussian-model-Shift+Width}
\end{equation}
as well as the dressed energy $E_{\text{d}}$ below threshold:
\begin{equation}
E_{\text{d}}=E_{\text{b}}+\Delta_{0}+\QDTalpha\kappa e^{\kappa^{2}\sigma^{2}}\left(1-\text{erf}(\kappa\sigma)\right)\label{eq:Gaussian-model-BoundState}
\end{equation}
where erf is the error function and $\kappa=\sqrt{2\mu(E_{\text{o}}-E_{\text{d}})}/\hbar$
is the binding wave number. In this model, the reduced width is given
by $\QDTalpha=\frac{2\mu}{4\pi\hbar^{2}}W_{0}^{2}$, and the zero-energy
shift by $\Delta_{0}=-\frac{\QDTalpha}{\sqrt{\pi}\sigma}$. Assume
that the scattering phase shift can be measured for different scattering
energies (as in high-energy experiments) or the dressed bound state
energy can be measured for different values of $E_{\text{b}}-E_{\text{o}}$
(as in ultracold-atom experiments). Then, fitting the data by Eqs.~(\ref{eq:Gaussian-model-Shift+Width})
or Eq.~(\ref{eq:Gaussian-model-BoundState}) should in general unambiguously
determine the parameters $\gamma$, $\sigma$ and $E_{\text{b}}-E_{\text{o}}$.

Figure\,\ref{fig:Gaussian-model} illustrates how different values
of the closed-channel parameter $\sigma$ at fixed $\QDTalpha$ lead
to different scattering phase shifts $\eta(E)$ and different dressed
bound-state energies $E_{\text{d}}$. For this particular model,
only three different measurements are required to determine the three
parameters of the model, enabling a characterisation of the bare bound
state.

This Gaussian model can be regarded as the regularised version of
a contact model corresponding to the leading order of the low-energy
effective field theory describing the resonance, as is often done
in the context of nuclear~\cite{Cohen2004} of hadron resonances~\cite{Kinugawa2024}.
After renormalisation, the parameter $\sigma$ can be set to arbitrarily
small values to recover the contact limit, yielding results that are
independent of $\sigma$. This $\sigma$-independent universal theory
is valid in a low-energy region (i.e. close to the threshold), which
can be seen in Fig.~\ref{fig:Gaussian-model} where the dressed bound
state energy curves for different values of $\sigma$ all coincide
with a universal curve shown in dashed grey --- we come back to this
point in Sec.~\ref{subsec:Application-to-low-energy}. Away from
this region, the effective field theory requires higher orders, which,
like the simple Gaussian model with finite $\sigma$, introduce parameters
characterising the closed channel. Again, the general conclusion holds
in this case: with enough experimental data, these closed-channel
parameters can in principle be determined.

\textcolor{red}{}

\section{Quantum defect theory regime\label{sec:Quantum-defect-regime}}

\subsection{Insensitivity to the closed channel}

It will now be shown that there is a regime where the details of
the closed channel are undetermined by experimental observations
of the scattering shift or binding energy. This situation arises for
systems in which the inter-channel coupling occurs around a distance
$r_{\text{w}}$ where the open-channel wave functions are energy independent.
This happens when the potential $V_{\text{o}}(r)$ in the open channel
has the form $E_{\text{o}}+V_{\text{tail}}(r)$ beyond a certain distance
$r_{0}$, where $V_{\text{tail}}(r)\xrightarrow[r\to\infty]{}0$ is
a potential tail that is independent of the value of the open-channel
scattering length $a_{\text{o}}$, which is set by the form of $V_{\text{o}}(r)$
at shorter distances $r\lesssim r_{0}$. If the tail is deep enough,
for a given energy $E$, there is a range of distances $r_{0}\lesssim r\ll r_{\text{tail}}(E)$
where the kinetic energy is negligible with respect to the potential,
namely $|E-E_{\text{o}}|\ll\vert V_{\text{tail}}(r_{\text{tail}})\vert$.
In that region, the open-channel wave functions are energy independent,
i.e. all proportional to the threshold solution at $E=E_{\text{o}}$.
If the coupling occurs in that region, it is well known that one can
employ the quantum defect theory (QDT)~\cite{Greene1982,Mies1984,Mies1984b,Mies2000,Raoult2004,Gao2005,Gao2009,Hanna2009,Gao2011,Ruzic2013,Jachymski2013,Chilcott2022}
to accurately describe the system for all the energies above and below
the threshold $E_{\text{o}}$ that are smaller than $\vert V_{\text{tail}}(r_{\text{w}})\vert$.
Although the usual treatment of QDT  makes use of the short-distance
K and Y matrices, here all quantities shall be expressed in terms
of observables such as $a_{\text{o}}$ and $\QDTalpha$. Doing so,
one obtains a ``renormalised'' formulation of QDT.

Above the threshold, one finds that the scattering phase shift Eq.~(\ref{eq:scatteringPhaseShift})
is given by the following expressions for the shift and width~\seeAppendix{\ref{subsec:Width-and-shift}}:
\begin{equation}
\boxed{\Delta(k)=\Delta_{0}+\QDTX(k)\frac{\Gamma(k)}{2}}\;;\;\boxed{\frac{\Gamma(k)}{2}=\QDTalpha\frac{k\left[\QDTZ(k)\right]^{-2}}{1+\left[\QDTX(k)\right]^{2}}}\label{eq:QDTShift+Width}
\end{equation}
with $\QDTX(k)\equiv\left[\tan\QDTetabar(k)\right]^{-1}-ka_{\text{o}}\left[\QDTZ(k)\right]^{-2}$,
where $\QDTetabar(k)$ and $\QDTZ(k)$ are two dimensionless functions
universally determined by the tail of $V_{\text{o}}$ ~\seeAppendix{\ref{subsec:Positive-energy}}.
 Physically, $\QDTetabar$ is the difference $\eta^{(\infty)}-\eta^{(0)}$,
where $\eta^{(a)}$ denotes the scattering phase shift for a potential
with tail $V_{\text{tail}}$ and scattering length $a$, and $\QDTZ(k)$
is the amplitude of its radial wave function $u_{\infty}^{(k)}$ at
infinite scattering length in the energy-independent region where
$u_{\infty}^{(k)}(r)=\QDTZ(k)\times u_{\infty}^{(0)}(r)$, with the
zero-energy solution $u_{\infty}^{(0)}$ normalised so that $u_{\infty}^{(0)}(r)\xrightarrow[r\to\infty]{}1$.

Below the threshold, one finds an even simpler result for the dressed
bound state energy:
\begin{equation}
\boxed{E_{\text{d}}=E_{\text{b}}+\Delta_{0}+\frac{\QDTalpha}{\QDTlambdakappa-a_{\text{o}}}}\label{eq:QDT_BoundState}
\end{equation}
where the function $\QDTlambdakappa$ is determined purely from the
tail of $V_{\text{o}}$~\seeAppendix{\ref{subsec:Negative-energy}}.
In fact, the energy $-\frac{\hbar^{2}\kappa^{2}}{2\mu}$ as a function
of $\QDTlambdakappa$ simply corresponds to the bound-state spectrum
for a potential with tail $V_{\text{tail}}$ as a function of its
scattering length.  It is quite remarkable that the mere knowledge
of the bare bound state spectrum for $V_{\text{o}}$ as a function
of its scattering length entirely determines the dressed bound state
spectrum through Eq.~(\ref{eq:QDT_BoundState}) once $\QDTalpha$,
$a_{\text{o}}$, and $\tilde{E}_{\text{b}}\equiv E_{\text{b}}+\Delta_{0}-E_{\text{o}}$
are known.

Equations~(\ref{eq:QDTShift+Width}) and (\ref{eq:QDT_BoundState})
constitute the first main result of this paper. They allow to determine
the two-body observables for all energies above and below the threshold
from only the three quantities $\QDTalpha$, $a_{\text{o}}$, and
$\tilde{E}_{\text{b}}$. Note that these quantities can be extracted
from the zero-energy scattering length Eq.~(\ref{eq:scattering-length}),
within the approximation $a_{\text{bg}}\approx a_{\text{o}}$. It
is therefore possible to determine a two-body observable (e.g. the
dressed bound state energy) from the knowledge of another observable
(e.g. the scattering length), without ever knowing the bare bound
state causing the resonance, nor its coupling to the open channel.

The crucial point leading to this result is that the zero-energy shift
$\Delta_{0}$ is taken apart and the width is expressed in terms
of the reduced width $\QDTalpha$. These are the only quantities
that depend explicitly upon the three closed-channel parameters $W_{0},a_{\text{c}},a_{\text{c}}^{\prime}$
through the expressions~\seeAppendix{\ref{subsec:Width-and-shift}},
\begin{equation}
\QDTalpha=\frac{2\mu}{4\pi\hbar^{2}}W_{0}^{2}\left(1-a_{\text{o}}/a_{\text{c}}\right)^{2}\quad;\quad\Delta_{0}=\QDTalpha\frac{a_{\text{o}}-a_{\text{c}}-a_{\text{c}}^{\prime}}{\left(a_{\text{o}}-a_{\text{c}}\right)^{2}}.\label{eq:QDT_resonance-strength+zero-energy-shift}
\end{equation}
Here, $W_{0}$ characterises the strength of the coupling between
the open channel and the bare bound state, while $a_{\text{c}}$ and
$a_{\text{c}}^{\prime}$ are two lengths characterising the closed
channel. Like scattering lengths, both $a_{\text{c}}$ and $a_{\text{c}}^{\prime}$
can be either positive or negative. The length $a_{c}$ was introduced
in Ref.~\cite{Naidon2019}, which focused on a specific regime in
which $a_{\text{c}}^{\prime}=0$ and $a_{\text{c}}$ sets the phase
of oscillations of the bare bound state wave function in the coupling
region. For this reason, $a_{\text{c}}$ was dubbed the ``closed-channel
scattering length'', by analogy with the open-channel scattering
length $a_{\text{o}}$ setting the phase of oscillations in the open
channel. Note however that in general $a_{\text{c}}^{\prime}\ne0$
and $a_{\text{c}}$ cannot always be interpreted as a scattering length
for the closed channel. 

In this renormalised formulation where $\Delta_{0}$ and $\QDTalpha$
are taken apart, one can now see the distinctive property of the QDT:
since the closed-channel parameters $W_{0},a_{\text{c}},a_{\text{c}}^{\prime}$
only affect the values of $\QDTalpha$ and $\Delta_{0}$, they cannot
be individually determined from the observables of Eq.~(\ref{eq:QDTShift+Width})
or (\ref{eq:QDT_BoundState}). The resonance is thus largely independent
of the details of the closed channel. This is in sharp contrast with
Eqs.~(\ref{eq:Gaussian-model-Shift+Width}-\ref{eq:Gaussian-model-BoundState}),
which depend explicitly on the closed-channel parameter $\sigma$
even after taking apart the zero-energy width $\Delta_{0}$ and the
reduced width $\QDTalpha$.

\begin{figure}[!t]
\includegraphics[viewport=0bp 5bp 188bp 210bp,clip,width=8.5cm]{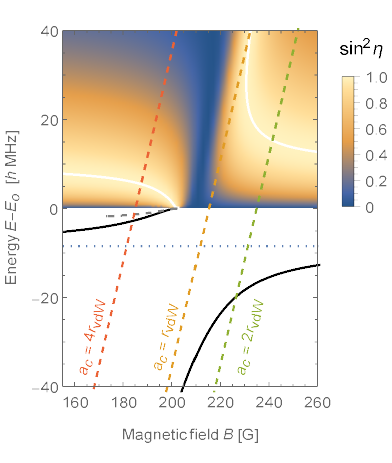}\caption{\label{fig:QDT-model}Energy spectrum and scattering phase shift of
the $^{40}$K $ab$ Feshbach resonance near $B_{0}=202$\,G. The
solid curve below the continuum threshold shows the dressed bound
state energy obtained from the QDT formula Eq.~(\ref{eq:QDT_BoundState}),
and the shading above the continuum threshold shows the quantity $\sin^{2}\eta$
obtained from Eq.~(\ref{eq:scatteringPhaseShift}) with the QDT formulas
Eq.~(\ref{eq:QDTShift+Width}). The resonance ($\sin^{2}\eta=1$)
is shown as a white curve. The functions $\QDTlambdakappa$, $\QDTetabar(k)$,
and $\QDTZk(k)$ are obtained for a van der Waals tail $V_{\text{tail}}(r)=-\frac{\hbar^{2}}{2\mu}\left(2r_{\text{vdW}}\right)^{4}/r^{6}$,
and the following parameters are used: $\tilde{E}_{\text{b}}=E_{\text{b}}+\Delta_{0}-E_{\text{o}}=\delta\mu\times(B-B_{0})$
with $\delta\mu/h=2.35$\,MHz/G, $\QDTalpha/h=50$\,MHz$\times r_{\text{vdW}}$,
$a_{\text{bg}}=2.635\thinspace r_{\text{vdW}}$, and $E_{\text{vdW}}/h=\frac{\hbar}{4\pi\mu r_{\text{vdW}}^{2}}=21\thinspace$MHz.
The horizontal dotted line shows the bound state energy of the open
channel. Note that the coupling of this bound state to the closed
channel creates an avoided crossing that splits the dressed bound
state energy curve into two branches: one on the left side which reaches
the open-channel threshold at $B_{0}$, and one on the right side
which asymptotes to the open-channel bound state energy. The dashed
grey curve shows the dressed bound state energy obtained from the
QDT Eq.~(\ref{eq:QDT_BoundState}) in the zero-range limit, corresponding
to Eq.~(\ref{eq:zeroRangeQDT2}). This plot reproduces Fig.\,13
of Ref.\,\cite{Chin2010}, except that the bare bound state energy
$E_{b}$ is shifted. The dashed slanted lines show the bare bound
state energy $E_{b}$ for three different models, whose closed-channel
parameters are (arbitrarily) set to $a_{\text{c}}^{\prime}=0$ and
$a_{\text{c}}=4r_{\text{vdW}},r_{\text{vdW}},2r_{\text{vdW}}$, respectively
from left to right, and $W_{0}$ is set to maintain the same value
of $\QDTalpha$. All models reproduce exactly the same dressed energy
and scattering phase shift. This shows that the position of the bare
bound state is arbitrary and not constrained by the observables.}
\end{figure}

\subsection{Application to magnetic Feshbach resonances\label{subsec:Application-to-magnetic}}

The QDT typically applies to ultracold atoms undergoing a magnetic
Feshbach resonance~\cite{Chin2010,Greene1982,Gao1998,Ruzic2013,Chilcott2022}.
The QDT regime is reached due to the deep van der Waals tail $V_{\text{tail}}(r)=-C_{6}/r^{6}=-\frac{\hbar^{2}}{2\mu}\left(2r_{\text{vdW}}\right)^{4}/r^{6}$
of the interatomic interactions, where $r_{\text{vdW}}$ is the van
der Waals length. In these systems, the bare bound energy $E_{\text{b}}$
in Eqs.~(\ref{eq:scatteringPhaseShift},\ref{eq:scattering-length},\ref{eq:QDT_BoundState})
is related to the magnetic field intensity $B$ by the Zeeman shift
through the relation $E_{\text{b}}+\Delta_{0}-E_{\text{o}}=\delta\mu\times(B-B_{0})$,
where $\delta\mu$ is the magnetic moment difference between the open
and closed channels, and $B_{0}$ is the magnetic field intensity
at which the resonance is observed at the threshold. The reduced width
is related to the observed magnetic width $\Delta B$ by $\QDTalpha=a_{\text{bg}}\delta\mu\times\Delta B$.
Thus, once the physical parameters $r_{\text{vdW}}$, $\delta\mu$,
$\Delta B$, $B_{0}$, and $a_{\text{bg}}\approx a_{\text{o}}$ are
known, all two-body observables can be determined from Eqs.~(\ref{eq:QDTShift+Width})
and (\ref{eq:QDT_BoundState}).

An example is shown in Fig.\,\ref{fig:QDT-model} for a resonance
between $^{40}$K atoms. It is described by three models with different
values of $W_{0},a_{\text{c}},a_{\text{c}}^{\prime}$, but conforming
to the same renormalised QDT given by Eqs.~(\ref{eq:QDTShift+Width})
and (\ref{eq:QDT_BoundState}). Thus there is no way of determining
the values of the closed-channel parameters from the observables shown
in that figure. Of course, if one could alter $a_{\text{o}}$ independently
of the other model parameters, then the values of $W_{0}$ and $a_{\text{c}}$
could be inferred from the change in $\QDTalpha$ by virtue of Eq.\,(\ref{eq:QDT_resonance-strength+zero-energy-shift})~\cite{Naidon2019}.
However this does not appear to be possible experimentally, and in
any case the value of $a_{\text{c}}^{\prime}$ would remain undetermined.
One must conclude that although the two-channel QDT provides an excellent
description of isolated resonances, its closed-channel parameters
$W_{0},a_{\text{c}},a_{\text{c}}^{\prime}$ are not fully constrained
by observables, and thus the shift $\Delta_{0}$ and the bare energy
$E_{\text{b}}$ are ambiguous quantities.

Here, the conservative point of view is taken that only scattering
phase shifts and bound state energies are fundamentally observable
at the two-body level. Other short-distance quantities can be observed
by involving a third body (such as a photon or another atom), as discussed
in Sec.~\ref{sec:Short-distance-physics}.

\subsection{Application to low-energy resonances (zero-range limit)\label{subsec:Application-to-low-energy}}

The QDT also applies to any resonance whose energy is very close
to the threshold. Indeed, for energies sufficiently close to the threshold,
the wave functions are energy-independent within the range of interactions,
because the potentials and couplings appear very deep compared with
the considered energies. The QDT formalism can therefore be applied,
and the energy-independent region can be approximated by a vanishingly
small region compared to the typical extent of wave functions. In
this limit, one obtains the analytic expressions~\seeAppendix{\ref{subsec:Case-of-contact}}
\begin{align}
A(k) & =1,\quad\bar{\eta}(k)=\pi/2,\label{eq:zeroRangeQDT1}\\
\lambda(\kappa) & =1/\kappa.\label{eq:zeroRangeQDT2}
\end{align}
This leads to a universal behaviour of near-threshold resonances
that is independent of the closed channel's details.

This zero-range QDT regime is nothing but the oft-used ``two-channel
zero-range model''~\cite{Lee1954,Petrov2004,Gogolin2008,Nishida2012}.
It is easy to check from Eq.~(\ref{eq:QDTShift+Width}) and (\ref{eq:scatteringPhaseShift})
that the effective range in this regime always has a negative value
--- \seeAppendix{\ref{subsec:Low-energy}}, 
\begin{equation}
r_{\text{eff}}=-2R_{\star}\left(1-\frac{a_{\text{bg}}}{a}\right)^{2}\label{eq:TwoChannelContactEffectiveRange}
\end{equation}
where the length $R_{\star}=\hbar^{2}/(2\mu\QDTalpha)$ characterises
the width of the resonance~\cite{Petrov2004}. This negative effective
range corresponds to a limit commonly called ``narrow'' or ``closed-channel
dominated'' Feshbach resonance~\cite{Chin2010} in the context of
cold atoms, and is obtained when $R_{\star}$ is much larger than
the range of interactions. Thus, if the resonance has in fact a positive
effective range, the two-channel zero-range universal regime only
applies at small energies where the effective range correction is
negligible. For those small energies, it reduces to the single-channel
zero-range universal regime that is parametrised by the scattering
length only. This zero-range universality is well known both in ultracold-atom
physics~\cite{Hammer2004} and hadron physics~\cite{Braaten2005,Hyodo2013}.

For instance, the zero-range universal limit can be seen in the case
of the magnetic Feshbach resonance of Fig.~\ref{fig:QDT-model}:
close to the threshold, the dressed bound state energy (solid black
curve) approaches the universal limit (dashed grey curve) obtained
from the QDT Eq.~(\ref{eq:QDT_BoundState}) with Eq.~(\ref{eq:zeroRangeQDT2}).
The zero-range universal limit can also be seen in the Gaussian model
of Fig\@.~\ref{fig:Gaussian-model}: as already mentioned in Sec.~\ref{sec:Generic-example},
close to the threshold, the curves for different values of $\sigma$
all coincide with the zero-range QDT (dashed grey curve) obtained
with Eqs.~(\ref{eq:zeroRangeQDT2}). In this low-energy limit, the
closed-channel parameter $\sigma$, and thus $\Delta_{0}$ and $E_{\text{b}}$,
become irrelevant, just as in Sec\@.~\ref{subsec:Application-to-magnetic}. 

However, away from the zero-range universal regime, there is a clear
discrepancy between magnetic Feshbach resonances and other kinds of
Feshbach resonances.

On the one hand, magnetic Feshbach resonances remain described by
the van der Waals QDT away from the threshold, since their interactions
feature a deep van der Waals tail. This results in a dressed bound
state (the single solid black curve of Fig.~\ref{fig:QDT-model})
that remains the same whatever the closed-channel parameters. 

On the other hand, resonances with shallow interactions are not described
by a QDT away from the threshold. Thus they become sensitive to the
closed channel details, as illustrated by the several curves of Fig\@.~\ref{fig:Gaussian-model}
obtained for different closed-channel parameters. This is the case,
for instance, for hadron resonances, since hadronic interactions feature
a shallow tail~\cite{Fabbietti2021}. Hence, the closed-channel parameters
of hadron resonances that are not very close to the threshold could
in principle be identified with enough data.

There have already been indications~\cite{Song2022,Albaladejo2022,Kinugawa2022a,Kinugawa2024}
that some hadron resonances significantly deviate from the zero-range
universal regime. For example, some resonances feature a positive
effective range, which by construction cannot be reproduced by Eq.~(\ref{eq:TwoChannelContactEffectiveRange}),
or lead to an open-channel fraction $X$ (also called ``compositeness''~\cite{Weinberg1963})
that exceeds unity when evaluated in the zero-range limit with $a_{\text{o}}=0$.
Table II in Ref.~\cite{Kinugawa2022a} lists several hadron resonances
with their corresponding binding energy, scattering length, and effective
range, obtained either from experimental data or ab initio calculations.
These three quantities cannot in general be reproduced by the zero-range
QDT with $a_{\text{o}}=0$, because there are only two parameters
in that theory, $\QDTalpha$ and $\tilde{E}_{\text{b}}$.

One can of course include a non-zero scattering length $a_{\text{o}}$
in the open channel. For instance, in the case of the X(3872) state~\cite{Choi2003},
suspected to result from a resonance between a pair of $D^{0}$ and
$\bar{D}^{*0}$ mesons and a compact $\bar{c}c$ bare bound state~\cite{Aaij2020},
fitting the quantities $E_{\text{d}}-E_{\text{o}}=-18$~keV, $a=28.5$~fm,
and $r_{\text{eff}}=-5.34$~fm listed in Ref.~\cite{Kinugawa2022a},
leads to $a_{\text{o}}=25.3$~fm. Since this model is in a QDT regime,
the fit does not provide any information about the bare bound state.

Eventually though, as more data is accumulated, it should prove impossible
to reproduce all data with only the three parameters of the zero-range
QDT, and models beyond it will become necessary. For example, one
may fit the above data with the nonrelativistic Gaussian model of
Eqs.~(\ref{eq:Gaussian-model-Shift+Width}-\ref{eq:Gaussian-model-BoundState}).
In this case, the extra parameter is given by the closed-channel parameter
$\sigma$, and one finds $\sigma=23.2$~fm. Since this model is not
in a QDT regime, it allows to determine the mass of the bare bound
state with respect to the $D^{0}$-$\bar{D}^{*0}$ threshold, namely
$E_{\text{b}}-E_{\text{o}}=-10.1$~keV. Of course, the significance
of this value is tied to one's trust in the model. The simplistic
Gaussian model is unlikely to provide an adequate description of the
X(3872) state, not to mention the complications related to the proximity
of other thresholds and decay channels~\cite{Aaij2020}. It nevertheless
illustrates how a model beyond the zero-range QDT regime can extract
some information about the compact core from experimental data.

\section{Discussion\label{sec:Discussion}}

\subsection{Closed-channel fraction}

Let us now mention a remarkable point. While the properties of the
bare bound state appear to be unobservable in the QDT regime, its
proportion $Z=1-X$ in the dressed bound state (called the ``closed-channel
fraction''~\cite{Timmermans1999,Partridge2005,Chen2005,Koehler2006,Zhang2009,Werner2009,Liu2021}
in the context of ultracold-atom physics and ``elementariness''~\cite{Weinberg1963,Hyodo2013,Li2022,Song2022,Albaladejo2022,Kinugawa2024}
in hadron physics) is observable and has indeed been measured in ultracold-atomic
systems~\cite{Partridge2005,Liu2021}. It can be easily calculated
from the Hellmann-Feynman theorem~\cite{Werner2009}, yielding $Z=dE_{\text{d}}/dE_{\text{b}}$.

Quite naturally, the closed-channel fraction in general depends on
the closed-channel details. For instance, for the Gaussian model,
taking the derivative of Eq.\,(\ref{eq:Gaussian-model-BoundState})
with respect to $E_{\text{d}}$ results in a closed-channel fraction
$Z$ that explicitly depends on the parameter $\sigma$ and thus
$\Delta_{0}$.

In contrast, in the QDT regime, taking the derivative of Eq.\,(\ref{eq:QDT_BoundState})
with respect to $E_{\text{d}}$ gives an expression that is independent
of the closed-channel parameters, and in particular of the shift $\Delta_{0}$.
It may be surprising that the fraction of the bare bound state in
the dressed wave function remains unaltered, even though the bare
bound state energy itself can be arbitrarily shifted away by $\Delta_{0}$.
For instance, one would intuitively think that the fraction goes to
unity only when the dressed energy approaches the bare energy. However,
the formula $Z=dE_{\text{d}}/dE_{\text{b}}$ shows that this is the
case even when the two energy curves are parallel to each other.
Physically, it means that even away from the resonance where the
dressed bound state is almost purely in the bare state, its energy
may be significantly shifted from the bare state energy through the
coupling to the open channel.  This reconciles the two facts that
the closed-channel bare bound state is not directly observable but
its fraction in the dressed bound state is.

Incidentally, one can also understand from these considerations that
the intuitive picture according to which the dressed bound state results
from an avoided crossing between the bare bound states in the open
and closed channels does not always hold. For instance, in Fig.~\ref{fig:QDT-model},
the dressed bound state (solid black curve) appears to result from
the avoided crossing between the open-channel bound state (dotted
line) and the bound state in the closed channel (orange dashed line)
corresponding to $a_{\text{c}}=r_{\text{vdW}}$. However, for the
other values of $a_{\text{c}}$ leading to different bare bound state
energies in the closed channel (dashed red or green curve), the avoided
crossing picture is much less apparent, even though the observables
remain the same. The reason is that an avoided crossing results from
the coupling of only two discrete bound states, whereas here the continuum
of states in the open channel can play a significant role and strongly
alter the avoided crossing picture.

\subsection{Dependence on the closed channel}

Even though quantities such as $E_{\text{b}}$ and $\Delta_{0}$
are found to be ambiguous and non-observable in the QDT regime, they
do have definite values for a given model, and these values depend
on the closed-channel parameters. In particular, the following expression
for the zero-energy shift $\Delta_{0}$~\cite{Goral2004,Julienne2006,Koehler2006,Nygaard2006,Chin2010,Jachymski2013,Kraats2023},
\begin{equation}
\frac{\QDTalpha}{\bar{a}}\frac{\frac{a_{\text{o}}}{\bar{a}}-1}{1+(\frac{a_{\text{o}}}{\bar{a}}-1)^{2}}\label{eq:WrongQDT_zero-energy-shift}
\end{equation}
has been shown to be incorrect in Ref.~\cite{Naidon2019}, resulting
from an invalid approximation in the QDT formalism. It can readily
be seen that this expression depends only on $a_{\text{o}}$ and
the characteristic range $\bar{a}$ of the open-channel potential
$V_{\text{o}}(r)$, but has no dependence on the closed channel, in
disagreement with the correct expression in Eq.\,(\ref{eq:QDT_resonance-strength+zero-energy-shift}).
Nevertheless, since $\Delta_{0}$ is unobservable, it is always possible
for fixed values of $\QDTalpha$ and $a_{\text{o}}$ to devise a model
with a choice of $W_{0},a_{\text{c}},a_{\text{c}}^{\prime}$ satisfying
Eq.\,(\ref{eq:WrongQDT_zero-energy-shift}), as done in Refs.~\cite{Goral2004,Kraats2023}.
This arbitrary choice does not affect the two-body observables. Thus,
while the value of $\Delta_{0}$ in Eq.\,(\ref{eq:WrongQDT_zero-energy-shift})
has no special significance, its use in these works has no consequence
on two-body observables. 

\begin{figure}
\includegraphics[width=8.5cm]{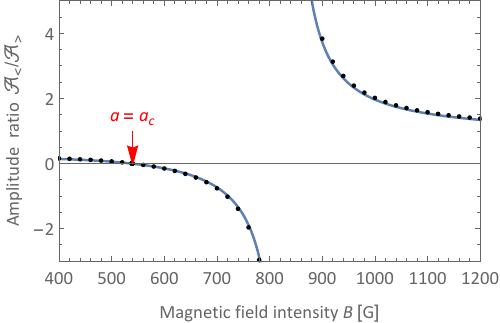}

\caption{\label{fig:Lithium6}Ratio of the amplitudes $\mathcal{A}_{<}$ and
$\mathcal{A}_{>}$ defined in Eq.~(\ref{eq:Short-range-Open-Channel})
as a function of magnetic field, for the $^{6}$Li $ab$ resonance
near $B_{0}=834$~G. The dots correspond to a realistic calculation
with five hyperfine channels, where the two amplitudes are obtained
from the triplet component of the zero-energy open-channel wave function.
The solid curve represents Eq.~(\ref{eq:Short-range-Amplitude-Ratio})
with the values $a_{\text{c}}=2.391\,$nm and $a_{\text{o}}=-112.8\,$nm,
corresponding respectively to the singlet and triplet scattering lengths.
The scattering length $a$  is given by Eq.~(\ref{eq:scattering-length})
with $a_{\text{bg}}=-84.89-24.19(B/B_{0}-1)+22.77(B/B_{0}-1)^{2}$
nm , $\gamma/h=62\,770\,$MHz~nm, and $\tilde{E}_{b}=E_{b}+\Delta_{0}-E_{\text{o}}=\delta\mu\times(B-B_{0})$
with $\delta\mu/h=2.8\,$MHz/G.}
\end{figure}
However, the works of Refs.~\cite{Goral2004,Kraats2023} are concerned
with three-body systems. This raises the important question whether
the value of $\Delta_{0}$, and more generally the closed-channel
parameters, could affect three-body observables. Indeed, three-body
observables are not only affected by two-body binding energies and
scattering phase shifts, but also off-the-energy-shell two-body quantities,
such as the short-distance two-body wave function. The next and final
section investigates how short-distance observables are affected by
the closed-channel parameters.

\section{Short-distance physics\label{sec:Short-distance-physics}}

The QDT gives a simple account of the short-distance two-body physics.
The wave function can be expressed by its open-channel radial component
$\uol(r)$ and closed-channel radial component $\ucl(r)$.  In the
energy independent region $r_{0}\lesssim r\ll r_{\text{tail}}(E)$,
one finds that $\uol(r)$ exhibits oscillations with a different phase
and amplitude beneath and beyond the coupling distance $r_{\text{w}}$
~\seeAppendix{\ref{subsec:Short-distance-amplitudes}}:
\begin{align}
\uol(r) & =\begin{cases}
\mathcal{A}_{<}\times u_{a_{\text{o}}}^{(0)}(r) & r\ll r_{\text{w}}\\
\mathcal{A}_{>}\times u_{a_{\text{eff}}}^{(0)}(r) & r\gg r_{\text{w}}
\end{cases}\label{eq:Short-range-Open-Channel}
\end{align}
where $u_{a}^{(0)}(r)\xrightarrow[r\to\infty]{}r-a$ is the zero-energy
solution of the open-channel potential with scattering length $a$,
and $a_{\text{eff}}$ is the energy-dependent scattering length $a_{\text{bg}}+\QDTalpha/(E-E_{b}-\Delta_{0})$.
This shows that the open-channel wave function has an unperturbed
form with amplitude $\mathcal{A}_{<}$ beneath the coupling region,
and a perturbed form with amplitude $\mathcal{A}_{>}$ beyond the
coupling region.

At low energy, the ratio of the amplitudes $\mathcal{A}_{<}/\mathcal{A}_{>}$
is given by:
\begin{equation}
\boxed{\frac{\mathcal{A}_{<}}{\mathcal{A}_{>}}=\frac{a-a_{\text{c}}}{a_{\text{o}}-a_{\text{c}}}}.\label{eq:Short-range-Amplitude-Ratio}
\end{equation}
This formula constitutes the second main result of this paper. It
gives a physical interpretation of $a_{\text{c}}$ as the scattering
length at which the short-distance amplitude $\text{\ensuremath{\mathcal{A}_{<}}}$
vanishes. 

This dependence of the short-range amplitudes on the closed-channel
scattering length $a_{\text{c}}$ is illustrated in Fig\@.~\ref{fig:Lithium6}
for the case of the 834~G magnetic Feshbach resonance between $^{6}$Li
atoms. This case is fortunate, since a two-channel model has been
clearly identified for this resonance as originating from the coupling
of a spin triplet open channel with a spin singlet bare bound state.
This suggests a photoassociation experiment to measure $a_{\text{c}}$:
by photoassociating $^{6}$Li atoms to an excited triplet bound state
with an extent smaller than $r_{\text{w}}$, one could measure $\mathcal{A}_{<}$
from the photoassociation signal, and determine $a_{\text{c}}$ from
the magnetic field at which $\mathcal{A}_{<}$ vanishes. For other
multi-channel resonances, however, it remains a challenge to identify
the effective two-channel model in general. \medskip{}

\section{Conclusion\label{sec:Conclusion}}

In summary, this work clarifies the role of closed-channel parameters
in Feshbach resonances. On the one hand, it is found that they affect
two-body observables (scattering phase shifts and binding energies)
in the general case, but not in the case of resonances involving deep
interaction potentials, such as magnetic Feshbach resonances between
ultracold atoms. Thus, the closed-channel parameters of magnetic
Feshbach resonances cannot be determined from these observables. This
is in sharp contrast with resonances involving shallow interaction
potentials, such as hadron resonances, for which this situation occurs
only close to the open-channel threshold.

On the other hand, one of the closed-channel parameters, called the
``closed-channel scattering length'', is found to affect short-distance
two-body physics. In ultracold-atomic systems, this parameter could
be determined by photoassociation, and should also affect three-body
observables, such as three-body recombination loss rates. The closed-channel
scattering length could thus play a role in the determination of the
three-body parameter characterising the Efimov spectrum of three-body
states near a magnetic Feshbach resonance~\cite{Naidon2017}, which
has been measured in various experiments, and for which a full theoretical
understanding is still in progress~\cite{Xie2020,Secker2021,Kraats2023,Yudkin2023,Kraats2023a}.

\vspace{1cm}
This work was supported by the JSPS Kakenhi grant No. JP23K03292.
The author is grateful to P.~S.~Julienne, E.~Tiesinga, L.~Pricoupenko,
M.~Raoult, S.~Kokkelmans, N.~Kjærgaard, M. Oka, S.~Endo, T.~Hyodo,
and T.~Kinugawa for stimulating discussions on this topic. The author
is especially thankful to S.~Endo for carefully checking the formulas
in this paper.

\clearpage\appendix
\setcounter{section}{0}
\renewcommand{\theequation}{\thesection.\arabic{equation}}
\setcounter{equation}{0}

\part*{Appendix}

\section{Two-channel model\label{sec:Two-channel-model}}

The Hamiltonian for a two-channel model of a two-particle system reads,
\begin{equation}
H=\left(\begin{array}{cc}
H_{\text{oo}} & H_{\text{oc}}\\
H_{\text{co}} & H_{\text{cc}}
\end{array}\right)\label{eq:Hamiltonian}
\end{equation}
where the open-channel Hamiltonian $H_{\text{oo}}$ and the closed-channel
Hamiltonian $H_{\text{cc}}$ are given by
\begin{align}
H_{\text{oo}} & =T+V_{\text{o}}\label{eq:Hoo}\\
H_{\text{cc}} & =T+V_{\text{c}}\label{eq:Hcc}
\end{align}
where $T$ is the relative kinetic operator, which for non-relativistic
systems is given by $\langle\bm{p}\vert T\vert\bm{q}\rangle=\frac{\hbar^{2}p^{2}}{2\mu}\delta^{3}(\bm{p}-\bm{q})$
where $\mu$ is the reduced mass of the particles. The open-channel
interaction potential $V_{\text{o}}$ asymptotes to a certain energy
threshold $E_{\text{o}}$ with a potential tail $V_{\text{tail}}$,
i.e. $V_{\text{o}}(r)\xrightarrow[r\to\infty]{}E_{\text{o}}+V_{\text{tail}}(r)$.
The closed-channel potential $V_{\text{c}}$ asymptotes to a certain
energy $E_{\text{c}}>E_{\text{o}}$.

The wave function $\phi$ of the system has two components, $\phio$
and $\phic$, respectively for the open and closed channels. At energy
$E$, they satisfy the coupled Schrödinger equations,
\begin{align}
\left(T+V_{\text{o}}-E\right)\vert\phio\rangle+H_{\text{oc}}\vert\phic\rangle & =0\label{eq:CoupledEquation1}\\
\left(T+V_{\text{c}}-E\right)\vert\phic\rangle+H_{\text{co}}\vert\phio\rangle & =0\label{eq:CoupledEquation2}
\end{align}

For energy $E<E_{\text{c}}$ (such that the second channel is indeed
closed), these equations lead to:
\begin{align}
\vert\phio\rangle & =\vert\phiobar\rangle+G_{\text{o}}^{+}H_{\text{oc}}\vert\phic\rangle\label{eq:Psi_o}\\
\vert\phic\rangle & =G_{\text{c}}H_{\text{co}}\vert\phio\rangle\label{eq:Psi_c}
\end{align}

where $G_{\text{o}}^{+}=(E+i0^{+}-T-V_{\text{o}})^{-1}$ and $G_{\text{c}}=(E-T-V_{\text{c}})^{-1}$
are the resolvents of the open and closed channels, and $\vert\phiobar\rangle$
is the scattering eigenstate of the open-channel Hamiltonian at energy
$E$ and scattering direction $\hat{k}$, normalised as $\langle\phiobar\vert\phiobarp\rangle=\delta(E-E^{\prime})\delta(\hat{k}-\hat{k}^{\prime})$.

\section{Two-channel isolated resonance theory\label{sec:Two-channel-isolated-resonance}}

\subsection{Definition of the resonance shift and width\label{subsec:Definition-of-the-shift-and-width}}

The closed-channel potential $V_{\text{c}}$ is assumed to support
a bound state $\vert\phib\rangle$ with energy $E_{\text{b}}$:
\begin{equation}
H_{\text{cc}}\vert\phib\rangle=E_{\text{b}}\vert\phib\rangle\label{eq:SchrodingerEquation-BoundState}
\end{equation}
It is normalised as $\langle\phi_{\text{b}}\vert\phi_{\text{b}}\rangle=1$.
In the isolated resonance approximation, only this bound state gives
a significant contribution to the resonance, so that one may write:
\begin{equation}
G_{\text{c}}=\frac{\vert\phib\rangle\langle\phib\vert}{E-E_{\text{b}}}+G_{\text{c}}^{\text{nr}}\label{eq:IsolatedResonanceApproximation}
\end{equation}
where the non-resonant part $G_{\text{c}}^{\text{nr}}$ only gives
a small contribution from the other states of the closed channel.
This leads to:
\begin{align}
\vert\phio\rangle & =\vert\phibg\rangle+\frac{G_{\text{o}}^{+}\vert W\rangle\langle W\vert\phibg\rangle}{E-E_{\text{b}}-\Delta^{+}}\label{eq:Psi_o-IsolatedResonance}\\
\vert\phic\rangle & =\vert\phib\rangle\frac{\langle W\vert\phibg\rangle}{E-E_{\text{b}}-\Delta^{+}}+G_{\text{c}}^{\text{nr}}H_{\text{co}}\vert\phio\rangle\label{eq:Psi_c-IsolatedResonance}
\end{align}
with the short-hand notations 
\begin{align}
\vert W\rangle & \equiv H_{\text{oc}}\vert\phib\rangle\label{eq:W}\\
\vert\phibg\rangle & \equiv\vert\phiobar\rangle+G_{\text{o}}^{+}H_{\text{oc}}G_{\text{c}}^{\text{nr}}H_{\text{co}}\vert\phio\rangle
\end{align}
and
\begin{equation}
\boxed{\Delta^{+}\equiv\langle W\vert G_{\text{o}}^{+}\vert W\rangle\equiv\Delta-i\frac{\Gamma}{2}}\label{eq:DeltaPlus}
\end{equation}
which defines the energy-dependent shift $\Delta(E)$ and width $\Gamma(E)$.

\subsection{Partial wave expansion\label{subsec:Partial-wave-expansion}}

Combining Eqs.~(\ref{eq:Psi_o}-\ref{eq:Psi_c}) gives a closed equation
on $\phio$:
\[
\vert\phio\rangle=\vert\phiobar\rangle+G_{\text{o}}^{+}H_{\text{oc}}G_{\text{c}}H_{\text{co}}\vert\phio\rangle
\]
Making the partial wave expansion along the direction $\hat{k}$ of
the incoming wave, 
\begin{align}
\langle\bm{r}\vert\phio\rangle & \equiv\sum_{\ell}\frac{\phiol(r)}{r}Y_{\ell0}(\hat{r})\label{eq:partial-wave-expansion-psi}\\
\langle\bm{r}\vert\phiobar\rangle & \equiv\sum_{\ell}\frac{\phiobarl(r)}{r}Y_{\ell0}(\hat{r})\label{eq:partial-wave-expansion-phi}\\
\langle\bm{r}\vert H_{\text{oc}}G_{\text{c}}H_{\text{co}}\vert\bm{r}^{\prime}\rangle & \equiv\sum_{\ell}\frac{H_{\ell}(r,r^{\prime})}{rr^{\prime}}Y_{\ell0}(\hat{r})Y_{\ell0}^{*}(\hat{r}^{\prime})\label{eq:partial-wave-expansion-H}
\end{align}
one finds for each partial wave $\ell$ the following complex radial
wave equation:
\begin{equation}
\phiol(r)\!=\!\phiobarl(r)\!+\!\!\int_{0}^{\infty}\!\!\!\!\!\!dr^{\prime}\QDTgkloplus(r,r^{\prime})\!\int_{0}^{\infty}\!\!\!\!\!\!dr^{\prime\prime}H_{\ell}(r^{\prime},r^{\prime\prime})\phiol(r^{\prime\prime})\label{eq:ComplexRadialEquation}
\end{equation}
where the retarded partial-wave Green's function $\QDTgkloplus$ is
given by
\begin{equation}
\QDTgkloplus(r,r^{\prime})=-\frac{2\mu}{\hbar^{2}k}\;\QDTuok(r_{<})\;\QDTvokplus(r_{>})\label{eq:partial-wave-Greens-function}
\end{equation}
with $k=\sqrt{2\mu(E-E_{\text{o}})}/\hbar$, $r_{>}=\max(r,r^{\prime})$
and $r_{<}=\min(r,r^{\prime})$. The two functions $\QDTuok$ and
$\QDTvokplus\equiv\QDTvok+i\,\QDTuok$ are two independent solutions
of the partial-wave radial equation:
\begin{equation}
\left(-\frac{d^{2}}{dr^{2}}+\frac{\ell(\ell+1)}{r^{2}}+\frac{2\mu}{\hbar^{2}}\left[V_{\text{o}}(r)-E_{\text{o}}\right]-k^{2}\right)u(r)=0,\label{eq:Schrodinger-equation-for-u}
\end{equation}
satisfying
\begin{align}
\QDTuok(r) & \xrightarrow[r\to\infty]{}\sin(kr-\ell\pi/2+\etaok)\label{eq:u-regular}\\
\QDTvok(r) & \xrightarrow[r\to\infty]{}\cos(kr-\ell\pi/2+\etaok)\label{eq:u-irregular}
\end{align}
where $\etaok$ is the $\ell$-wave scattering phase shift of the
open channel. The solution $\QDTuok(r)$ is regular (vanishing when
$r\to0$), whereas the solutions $\QDTvokplus(r)$ and $\QDTvok(r)$
are irregular (non-vanishing for $r\to0$).

In the following, the notations $(A\vert B)\equiv\int drA(r)B(r)$
and $(A\vert B\vert C)\equiv\int dr\int dr^{\prime}A(r)B(r,r^{\prime})C(r^{\prime})$
will used.

From the definitions of $\phiobar$ and $\QDTuok$, one finds 
\begin{equation}
\phiobarl(r)=\bar{\mathcal{N}}_{\ell}\times\QDTuok(r)\label{eq:relation-between-u-and-phi}
\end{equation}
with the complex coefficient $\bar{\mathcal{N}}_{\ell}\equiv\sqrt{\frac{2\mu(2\ell+1)}{\pi\hbar^{2}k}}i^{\ell}e^{i\etaok}$.
The complex equation Eq.~(\ref{eq:ComplexRadialEquation}) can then
be made real by splitting the real and imaginary parts of the Green's
function Eq.~(\ref{eq:partial-wave-Greens-function}), and setting
\begin{equation}
\phiol(r)\equiv\mathcal{N}_{\ell}\times\uol(r)\label{eq:realRadialWavefunction}
\end{equation}
with the complex coefficient $\mathcal{N}_{\ell}\equiv\bar{\mathcal{N}}_{\ell}\left(1+i\frac{2\mu}{\hbar^{2}k}(\QDTuok|H_{\ell}|\uol)\right)^{-1}$.
This yields the following equation for the real radial wave function
$\uol$:
\begin{equation}
\boxed{\uol(r)=\QDTuok(r)\!+\!\int_{0}^{\infty}\!\!\!dr^{\prime}\QDTgklo(r,r^{\prime})\int_{0}^{\infty}\!\!\!dr^{\prime\prime}H_{\ell}(r^{\prime},r^{\prime\prime})\uol(r^{\prime\prime})}.\label{eq:RealRadialEquation}
\end{equation}
with the non-retarded partial-wave Green's function,
\begin{equation}
\QDTgklo(r,r^{\prime})\equiv-\frac{2\mu}{\hbar^{2}k}\;\QDTuok(r_{<})\;\QDTvok(r_{>})\label{eq:non-retarded-partial-wave-Green-s-function}
\end{equation}

\subsection{Isolated resonance\label{subsec:Isolated-resonance}}

\subsubsection{Scattering phase shift}

Using the isolated resonance decomposition Eq.~(\ref{eq:IsolatedResonanceApproximation})
in Eq.~(\ref{eq:RealRadialEquation}), and assuming that $\vert W\rangle=H_{\text{oc}}\vert\phib\rangle$
is of the form
\begin{equation}
\langle\bm{r}\vert W\rangle=\frac{w(r)}{r}Y_{\ell0}(\hat{r})\label{eq:SinglePartialWaveApproximation}
\end{equation}
 acting on a specific partial wave $\ell$, one obtains for that
partial wave:
\begin{equation}
H_{\ell}(r,r^{\prime})=\frac{w(r)w(r^{\prime})}{E-E_{b}}+H_{\ell}^{\text{nr}}(r,r^{\prime})\label{eq:H_ell}
\end{equation}
where $H_{\ell}^{\text{nr}}$ correspond to the non-resonant part
$H_{\text{oc}}G_{\text{c}}^{\text{nr}}H_{\text{co}}$. This gives
\begin{equation}
\uol(r)=\ubg(r)+(w\vert\uol)\frac{\int_{0}^{\infty}dr^{\prime}\QDTgklo(r,r^{\prime})w(r^{\prime})}{E-E_{b}}\label{eq:RadialEquation1}
\end{equation}
with the background function
\begin{equation}
\ubg(r)\equiv\QDTuok(r)+\int_{0}^{\infty}dr^{\prime}\QDTgklo(r,r^{\prime})w_{\text{nr}}(r^{\prime})\label{eq:ubg}
\end{equation}
where 
\begin{equation}
w_{\text{nr}}(r)=\int_{0}^{\infty}dr^{\prime}H_{\ell}^{\text{nr}}(r,r^{\prime})\uol(r^{\prime})\label{eq:w_nr}
\end{equation}
corresponds to the coupling to other states than the bare bound state
causing the resonance.

Applying $(w\vert$ to the left of Eq.~(\ref{eq:RadialEquation1})
to find $(w\vert\uol)$, and inserting the result back into Eq.~(\ref{eq:RadialEquation1})
gives
\begin{equation}
\boxed{\uol(r)=\ubg(r)+(w\vert\ubg)\frac{\int_{0}^{\infty}dr^{\prime}\QDTgklo(r,r^{\prime})w(r^{\prime})}{E-E_{\text{b}}-\Delta}}\label{eq:RadialEquation2}
\end{equation}
with the shift $\Delta=(w\vert\QDTgklo\vert w)$. At large distances,
the radial wave function becomes

\begin{equation}
\uol(r)=\QDTuok(r)-\left[\xi_{\text{nr}}+\frac{(\Gamma+\Gamma_{\text{nr}})/2}{E-E_{\text{b}}-\Delta}\right]\QDTvok(r)\label{eq:uo-large-distances}
\end{equation}
with the width $\Gamma/2=\frac{2\mu}{\hbar^{2}k}\left|(w\vert\QDTuok)\right|^{2}$
and the non-resonant corrections
\begin{align}
\Gamma_{\text{nr}}/2 & \equiv\frac{2\mu}{\hbar^{2}k}(w\vert\QDTgklo\vert w_{\text{nr}})(\QDTuok\vert w)\label{eq:Gamma_nr}\\
\xi_{\text{nr}} & \equiv\frac{2\mu}{\hbar^{2}k}(\QDTuok\vert w_{\text{nr}})\label{eq:xi_nr}
\end{align}

Using the asymptotic behaviours of $\QDTuok$ and $\QDTvok$ given
in Eqs.~(\ref{eq:u-regular}-\ref{eq:u-irregular}), one obtains
from Eq.~(\ref{eq:uo-large-distances}),
\begin{equation}
\uol(r)\xrightarrow[r\to\infty]{}\propto\sin\left(kr-\ell\pi/2+\eta\right)\label{eq:uol-asymptotic}
\end{equation}
with the scattering phase shift,
\begin{equation}
\etak=\etaok-\arctan\left(\xi_{\text{nr}}+\frac{(\Gamma+\Gamma_{\text{nr}})/2}{E-E_{\text{b}}-\Delta}\right)\label{eq:phase-shift0}
\end{equation}
Treating the non-resonant corrections as a first-order perturbation,
one finally arrives at
\begin{equation}
\boxed{\etak=\etabgk-\arctan\frac{\tilde{\Gamma}/2}{E-E_{\text{b}}-\Delta}}\label{eq:phase-shift}
\end{equation}
with the background phase shift:
\begin{equation}
\etabgk\equiv\etaok-\xi_{\text{nr}}\left[1+\left(\frac{\Gamma/2}{E-E_{\text{b}}-\Delta}\right)^{2}\right]^{-1}\label{eq:eta_bg}
\end{equation}
and the corrected width:

\begin{equation}
\tilde{\Gamma}\equiv\Gamma+\Gamma_{\text{nr}}\label{eq:Gamma_tilde}
\end{equation}
In the fully isolated resonance approximation, one neglects the non-resonant
corrections, yielding $\etabgk\approx\etaok$ and $\tilde{\Gamma}\approx\Gamma$
in Eq.~(\ref{eq:phase-shift}).

\subsubsection{Low-energy limit in the s wave}

In the case of s wave ($\ell=0$), the quantities $\Gamma,\Gamma_{\text{nr}}$,
and $\xi_{\text{nr}}$ for small $k$ are proportional to $k$ (being
proportional to $\QDTuok$) and thus one obtains the s-wave scattering
length:
\begin{equation}
\boxed{a=-\lim_{k\to0}\etak/k=a_{\text{bg}}-\frac{\tilde{\gamma}}{E_{b}+\Delta_{0}-E_{\text{o}}}}\label{eq:PhysicalScatteringLength}
\end{equation}
where $a_{\text{bg}}=-\lim_{k\to0}\etabgk/k=a_{\text{o}}+a_{\text{nr}}$
with $a_{\text{nr}}=\lim_{k\to0}\xi_{\text{nr}}/k$, and $\tilde{\gamma}=\lim_{k\to0}\tilde{\Gamma}/2k=\gamma+\gamma_{\text{nr}}$,
with $\gamma_{\text{nr}}\equiv\lim_{k\to0}\Gamma_{\text{nr}}/2k$.

One can more generally assume that:
\begin{align}
\Gamma_{k}/2 & =\gamma k\left(1+\beta k^{2}\right)+O(k^{3})\label{eq:Gamma-low-energy}\\
\tilde{\Gamma}_{k}/2 & =\tilde{\gamma}k\left(1+\tilde{\beta}k^{2}\right)+O(k^{3})\label{eq:Gamma-tilde-low-energy}\\
\Delta_{k} & =\Delta_{0}+\alpha k^{2}+O(k^{3})\label{eq:Delta-low-energy}\\
\frac{k}{\tan\eta_{\text{bg}}} & =-\frac{1}{a_{\text{bg}}}+\frac{1}{2}r_{\text{bg}}k^{2}+O(k^{3})\label{eq:delta-bg-low-energy}
\end{align}
so that one finds from Eq.~(\ref{eq:phase-shift}) the following
low-energy expansion:
\begin{equation}
\frac{k}{\tan\eta}=-\frac{1}{a}+\frac{1}{2}r_{\text{eff}}k^{2}+O(k^{3})\label{eq:general-effective-range-expansion}
\end{equation}
with $a$ given by Eq.~(\ref{eq:PhysicalScatteringLength}) and the
effective range $r_{\text{eff}}$ given by:
\begin{equation}
\boxed{r_{\text{eff}}=2\left(\frac{\alpha}{\tilde{\gamma}}-R_{\star}+\frac{aa_{\text{bg}}+\tilde{\beta}}{a-a_{\text{bg}}}\right)\left(1-\frac{a_{\text{bg}}}{a}\right)^{2}+r_{\text{bg}}\frac{a_{\text{bg}}^{2}}{a^{2}}}\label{eq:general-effective-range}
\end{equation}
where the following length~\cite{Petrov2004} is introduced:
\begin{equation}
R_{\star}\equiv\frac{\hbar^{2}}{2\mu\tilde{\gamma}}\label{eq:Rstar-definition}
\end{equation}
Close to the resonance ($a\to\infty$), the effective range reduces
to:
\begin{equation}
r_{\text{eff}}=2\left(\frac{\alpha}{\tilde{\gamma}}+a_{\text{bg}}-R_{\star}\right)\label{eq:general-effective-range-at-unitarity}
\end{equation}

\section{Quantum Defect Theory\label{sec:Quantum-Defect-Theory}}

\begin{figure*}[!t]
\hfill{}\includegraphics[width=8cm]{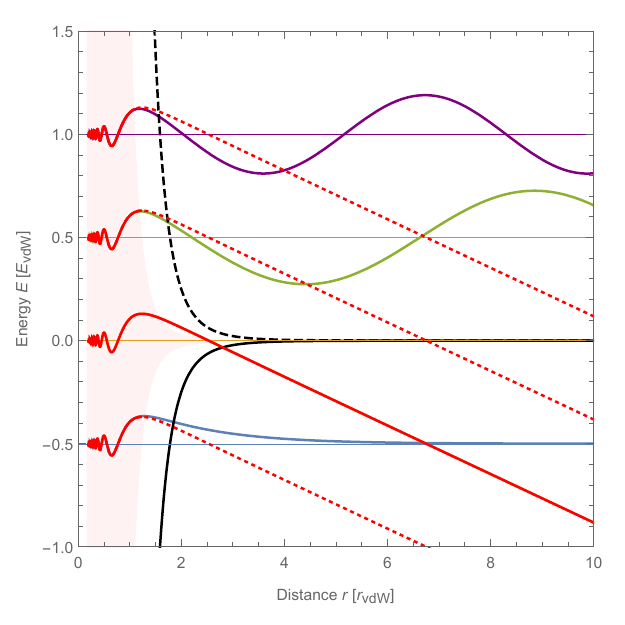}\hfill{}\includegraphics[width=8cm]{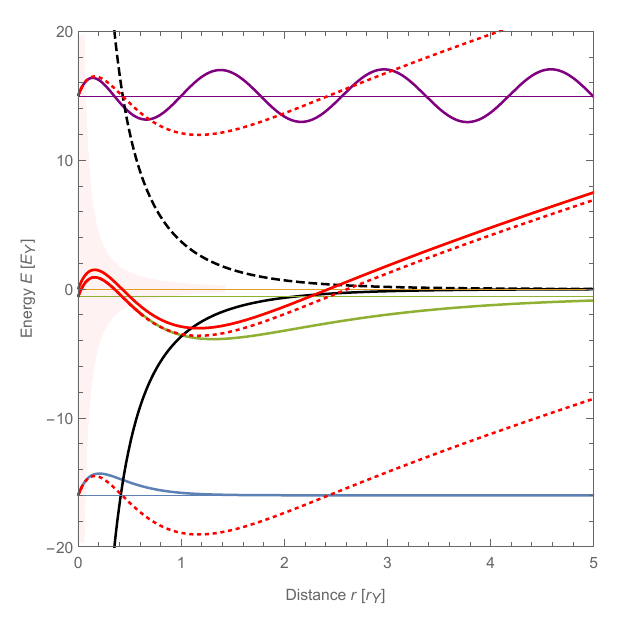}\hfill{}

\caption{\label{fig:Energy-independence}Energy independence of s-wave radial
wave functions for two potentials: a deep van der Waals potential
$V(r)=-E_{\text{vdW}}\left(\frac{r_{\text{vdW}}}{r}\right)^{6}$ with
many bound states (left panel) and a shallow Yukawa potential $V(r)=-10E_{\text{Y}}\frac{\exp(-r/r_{\text{Y}})}{r/r_{\text{Y}}}$
with only two bound states (right panel). Here, $r_{\text{vdW}}$
and $r_{\text{Y}}$ are the van der Waals and Yukawa ranges, and $E_{\text{vdW}}\equiv\frac{\hbar^{2}}{2\mu r_{\text{vdW}}^{2}}$
and $E_{\text{Y}}\equiv\frac{\hbar^{2}}{2\mu r_{\text{Y}}^{2}}$ are
their associated energies. Each panel shows $V(r)$ (black curve)
and $-V(r)$ (black dashed curve), along with the radial wave functions
at selected energies (the curves are shifted according to their respective
energies). The zero-energy radial wave function is shown in solid
red, and superimposed as a red dotted curve onto the curves corresponding
to radial wave functions at other energies. The region of energy independence
$r_{0}\lesssim r\ll r_{\text{tail}}$ (or equivalently $r\gtrsim r_{0}$
and $E\ll|V(r)|$) is shown as a pink shaded area. In that region,
the zero-energy wave function matches the wave functions at other
energies, and is shown in solid red. The energy independent region
is much smaller in the case of a shallow potential, for which the
quantum defect theory is barely applicable, except at very small energies
$\ll E_{\text{Y}}$ and large distances $\gg r_{\text{Y}}$ corresponding
to the zero-range limit. For instance, the last bound state is well
described by the zero-range limit for $r\gg r_{\text{Y}}$, because
it can be determined from a boundary condition at $r\approx r_{\text{Y}}$
that is similar to that of the zero-energy state. That is not the
case for the lowest bound state, which differs too much from the zero-energy
state.}

\end{figure*}
The key point of the quantum defect theory is that when an interaction
potential is sufficiently deep in a certain region $r_{0}\lesssim r\ll r_{\text{tail}}$,
the wave functions in that region are nearly energy-independent for
a range of energies that remain much smaller than the potential energy
$V(r_{\text{tail}})$. In that region (and only in that region), the
wave function at any of these energies is accurately described by
a superposition of two independent solutions of the potential at zero
energy. For a specific choice of two reference solutions, there is
a particular linear combination reproducing the wave function in the
region. The coefficients of this linear combination can be parametrised
by a global normalisation factor, and a parameter called the \emph{quantum
defect}. 

In the following, the reference functions are chosen as the two s-wave
radial solutions of the potential at zero-energy, with respectively
zero and infinite scattering length:
\begin{align}
\uzero(r) & \xrightarrow[r\to\infty]{}r\label{eq:f0asymptote}\\
\uinfty(r) & \xrightarrow[r\to\infty]{}1\label{eq:finfasymptote}
\end{align}
The zero-energy solution with scattering length $a$ is thus $\uzero(r)-a\uinfty(r)\xrightarrow[r\to\infty]{}r-a$.
With this choice, the quantum defect is simply the s-wave scattering
length $a$.

For example, for a van der Waals interaction $V(r)\to-C_{6}/r^{6}$,
one has: 
\begin{align}
\uzero(r) & =r_{\text{vdW}}\sqrt{x}\Gamma(3/4)J_{-1/4}(2x^{-2})\label{eq:van-der-Waals-f0}\\
\uinfty(r) & =\sqrt{x}\Gamma(5/4)J_{1/4}(2x^{-2})\label{eq:van-der-Waals-finfinity}
\end{align}
where $x=r/r_{\text{vdW}}$ and $r_{\text{vdW}}$ is the van der
Waals length $r_{\text{vdW}}\equiv\frac{1}{2}\left(2\mu C_{6}/\hbar^{2}\right)^{1/4}$.
The range of energy-independence at energy $\vert E\vert=\hbar^{2}k^{2}/2\mu$
is given by $r_{0}\lesssim r\ll r_{\text{tail}}$ with $r_{\text{tail}}=r_{\text{vdW}}^{2/3}k^{-1/3}$.
It is illustrated in the left panel of Fig.~\ref{fig:Energy-independence}
as a pink shaded area.

Interestingly, the quantum defect approach also applies to contact
interactions. In this case, the region of energy-independence is restricted
to the neighbourhood of $r=0$ (i.e. $r_{0}=r_{\text{tail}}=0$) but
extends to any energy. The two reference solutions are simply $\uzero(r)=r$
and $\uinfty(r)=1$. This is of course an idealisation, which can
be regarded as the limit of a short-range interaction potential with
vanishing range and infinite depth. Physically, it describes the wave
functions of a short-range interaction potential for energies much
smaller than the potential depth and distances larger than the potential
range. The energy independent region in this case corresponds to energies
smaller than the potential depth and distances smaller than the potential
range. This is illustrated in the right panel of Fig.~\ref{fig:Energy-independence}
for a shallow Yukawa potential. In the contact interaction limit,
this region reduces to a boundary condition on the logarithmic derivative
at $r=0$. 

\subsection{Positive energy\label{subsec:Positive-energy}}

Let us now consider a potential $V(r)\xrightarrow[r\to\infty]{}0$
of s-wave scattering length $\QDTa$ and its regular and irregular
radial solutions $\QDTuk$ and $\QDTvk$ in the $\ell$th partial
wave at finite positive energy $E=\hbar^{2}k^{2}/2\mu>0$. The regular
function $\QDTuk$ is defined such that $\QDTuk(0)=0$, which gives
at large distance $\QDTuk(r)\to\sin(kr+\QDTetakla-\ell\pi/2)$ where
$\QDTetakla$ is the scattering phase shift. The irregular solution
$\QDTvk$ is chosen such that its phase at large distances is shifted
by $\pi/2$ with respect to $\QDTuk$.
\begin{align}
\QDTuk(r) & \xrightarrow[r\to\infty]{}\sin(kr+\QDTetakla-\ell\pi/2)\label{eq:ukasymptote}\\
\QDTvk(r) & \xrightarrow[r\to\infty]{}\cos(kr+\QDTetakla-\ell\pi/2)\label{eq:vkasymptote}
\end{align}

According to the quantum defect assumption, in the energy independent
region $r_{0}\lesssim r\ll r_{\text{tail}}$ the two functions $\QDTuk$
and $\QDTvk$ are linear combinations of the two zero-energy reference
solutions $\uzero$ and $\uinfty$ . The regular solution $\QDTuk$
is simply proportional to the zero-energy solution $\uzero-\QDTa\uinfty$
with scattering length $\QDTa$:
\begin{equation}
\QDTuk(r)\xrightarrow[r_{0}\lesssim r\ll r_{\text{tail}}]{}\QDTDkla(k)\left(\uzero(r)-\QDTa\uinfty(r)\right)\label{eq:Expansionuk}
\end{equation}
Similarly, the irregular solution $\QDTvk$ has the form:
\begin{equation}
\QDTvk(r)\xrightarrow[r_{0}\lesssim r\ll r_{\text{tail}}]{}\QDTPkla(k)\left(\uzero(r)-\QDTbkla(k)\uinfty(r)\right)\label{eq:Expansionvk}
\end{equation}

The Wronskian $W[\QDTuk,\QDTvk]=\QDTuk(\QDTvk)^{\prime}-(\QDTuk)^{\prime}\QDTvk$
has the conserved value $-k$ calculated from Eqs.\,(\ref{eq:ukasymptote}-\ref{eq:vkasymptote}),
so from the expressions of Eqs.\,(\ref{eq:Expansionuk}-\ref{eq:Expansionvk}),
one finds:
\begin{equation}
\boxed{\left[\QDTbkla(k)-\QDTa\right]\QDTDkla(k)\QDTPkla(k)=-k}\label{eq:RelationBetween_Dk_bk_a_Pk}
\end{equation}
which shows that only two of the functions $\QDTDkla,\QDTbkla,\QDTPkla$
are independent for a given $\QDTa$.

One can determine $\QDTetakla,\QDTDkla,\QDTbkla$, and $\QDTPkla$
for any scattering length $\QDTa$, by just knowing four functions
of $k$: $\QDTetazero,\QDTetainf,\QDTDklzero,\QDTZk$.
\begin{align}
\tan\QDTetakla & =\frac{\left(\QDTDklzero\right)^{-1}\sin\QDTetazero-\QDTa\left(\QDTZk\right)^{-1}\sin\QDTetainf}{\left(\QDTDklzero\right)^{-1}\cos\QDTetazero-\QDTa\left(\QDTZk\right)^{-1}\cos\QDTetainf}\label{eq:tandeltak-expression}\\
\QDTDkla & =\left[\left(\QDTDklzero\right)^{-2}+\left(\frac{\QDTa}{\QDTZk}\right)^{2}-2\QDTa\frac{\cos\QDTetabar}{\QDTDklzero\QDTZk}\right]^{-1/2}\label{eq:Dk-expression}\\
\QDTbkla & =\frac{\left(\QDTZk/\QDTDklzero\right)-\QDTa\cos\QDTetabar}{\cos\QDTetabar-\QDTa\left(\QDTDklzero/\QDTZk\right)}\label{eq:bk-expression}\\
\QDTPkla & =-\frac{\cos\QDTetabar-\QDTa\left(\QDTDklzero/\QDTZk\right)}{\sin\QDTetabar}\QDTDkla\label{eq:Pk-expression}
\end{align}
with the notations
\begin{equation}
\QDTetabarkla\equiv\QDTetainf-\QDTetazero\label{eq:etabar}
\end{equation}
and
\begin{align}
\QDTetazero & \equiv\lim_{\QDTa\to0}\QDTetakla\qquad;\qquad\QDTDklzero\equiv\lim_{\QDTa\to0}\QDTDkla\label{eq:delta0-Dk0-definitions}\\
\QDTetainf & \equiv\lim_{\QDTa\to-\infty}\QDTetakla\qquad;\qquad\QDTZk\equiv\lim_{\QDTa\to-\infty}-\QDTa\QDTDkla\label{eq:delta-infinity-Zk-definitions}
\end{align}

Again the four functions $\QDTetazero,\QDTetainf,\QDTDklzero,\QDTZk$
are not independent, because the Wronskian $W[\QDTukZero,\QDTukInfi]$
can be expressed at short distance as
\begin{equation}
W[\QDTDklzero\uzero,\QDTZk\uinfty]=\QDTDklzero\QDTZk\underbrace{W[\uzero,\uinfty]}_{-1}=-\QDTDklzero\QDTZk\label{eq:ShortRangeWronskian}
\end{equation}
and at large distance as:
\begin{align}
 & W[\sin(kr-\ell\pi/2+\QDTetazero),\sin(kr-\ell\pi/2+\QDTetainf)]\label{eq:LongRangeWronskian}\\
 & =-k\sin\left(\QDTetainf-\QDTetazero\right)\nonumber 
\end{align}
leading to the relation,
\begin{equation}
\boxed{\QDTDklzero(k)\QDTZk(k)=k\sin\QDTetabarkla(k)}\label{eq:RelationBetween_D0k_Zk_eta0_etainf}
\end{equation}

Using this relation, one can express $\QDTDkla,\QDTbkla,\QDTPkla$
in terms of only two functions $\QDTZk$ and $\QDTetabarkla$:
\begin{align}
\QDTDkla & =\frac{k}{\QDTZk}\left[1+\left(\QDTXkla\right)^{2}\right]^{-1/2}\label{eq:Dkla-in-terms-of-Z-eta}\\
\QDTbkla & =\frac{\left(\QDTZk\right)^{2}}{k}\left[\frac{1}{\QDTXkla}+\frac{1}{\tan\QDTetabarkla}\right]\label{eq:bkla-in-terms-of-Z-eta}\\
\QDTPkla & =-\QDTXkla\QDTDkla\label{eq:Pkla-in-terms-of-Z-eta}
\end{align}
where
\begin{equation}
\QDTXkla\equiv\frac{1}{\tan\QDTetabarkla}-\frac{k\QDTa}{\left(\QDTZk\right)^{2}}\label{eq:Xkla-definition}
\end{equation}

\subsection{Alternative choice\label{subsec:Alternative-choice}}

One may consider an alternative choice $\QDTvkplus(r)\equiv\QDTvk(r)+i\QDTuk(r)$
for the irregular function, that has the complex asymptote:
\begin{equation}
\QDTvkplus(r)\xrightarrow[r\to\infty]{}e^{i(kr+\QDTetakla-\ell\pi/2)}\label{eq:vk+asymptote}
\end{equation}
It can be expanded on $\uzero$ and $\uinfty$
\begin{equation}
\QDTvkplus(r)\xrightarrow[r_{0}\lesssim r\ll r_{\text{tail}}]{}\QDTPklaplus\left(\uzero(r)-\QDTbklaplus\uinfty(r)\right)\label{eq:Expansionvkplus}
\end{equation}
where the complex quantities $\QDTPklaplus$ and $\QDTbklaplus$ are
readily obtained from Eqs.~(\ref{eq:Expansionuk}-\ref{eq:Expansionvk}):
\begin{align}
\QDTPklaplus & =\QDTPkla+i\QDTDkla\label{eq:Pkplus}\\
\QDTbklaplus & =\frac{\QDTPkla\QDTbkla+i\QDTDkla\QDTa}{\QDTPkla+i\QDTDkla}\label{eq:bkplus}
\end{align}

The interest of this alternative choice is that the quantity $\QDTbklaplus$
is independent of $\QDTa$. Indeed, using Eqs.~(\ref{eq:Dk-expression}-\ref{eq:Pk-expression}),
one finds:
\begin{equation}
\boxed{\QDTbklplus=\left(\QDTZk/\QDTDklzero\right)e^{i\QDTetabarkla}}\label{eq:bkplus2}
\end{equation}
where the label $a$ is now dropped, due to the independence on $a$.

Again, from the Wronskian $W[\QDTuk,\QDTvk]=-k$, one finds
\begin{equation}
\boxed{\left(\QDTbklplus-\QDTa\right)\QDTDkla\QDTPklaplus=-k}\label{eq:RelationBetween_Dk_bkplus_a_Pkplus}
\end{equation}
From Eqs.~(\ref{eq:Pkplus}-\ref{eq:RelationBetween_Dk_bkplus_a_Pkplus})
one also finds the useful relations:
\begin{equation}
\frac{1}{\QDTbklplus-\QDTa}=\frac{1}{\QDTbkla-\QDTa}-\frac{i\left(\QDTDkla\right)^{2}}{k}=-\frac{\QDTDkla(\QDTPkla+i\QDTDkla)}{k}\label{eq:useful-relations}
\end{equation}

\subsection{Negative energy\label{subsec:Negative-energy}}

For negative energies $E=-\frac{\hbar^{2}\kappa^{2}}{2\mu}$ obtained
when $k$ is continued to imaginary values $i\kappa$, the quantity
$\QDTbklplus$ becomes real. For convenience, $b^{+}(i\kappa)$ is
denoted as $\QDTlambdakappa$. One can see from Eq.~(\ref{eq:vk+asymptote})
that for imaginary $k=i\kappa$, the irregular function $\QDTvkplus$
is exponentially decreasing at large distance. Equation~(\ref{eq:Expansionvkplus})
shows that if $\QDTlambdakappa$ happens to be equal to $a$, then
$\QDTvkplus$ is proportional to the regular solution $\QDTuk$, as
seen from Eq.~(\ref{eq:ukasymptote}). In this case, being both regular
at the origin and at infinity, the solution corresponds to a bound
state. This shows that $\QDTlambdakappa$ is simply the s-wave scattering
length $\QDTa$ of the potential at which there is a bound state in
the $\ell$th partial wave with energy $-\frac{\hbar^{2}\kappa^{2}}{2\mu}$.

\subsection{Calculation of the universal functions\label{subsec:Calculation-of-the-universal-functions}}

\subsubsection{General case}

The functions $\QDTetazero,\QDTetainf,\QDTDklzero,\QDTZk$ may in
some cases be calculated analytically for a given tail of the potential
$V$, for example in the case of a contact interaction (see below).
If only the analytical forms of $\uzero$ and $\uinfty$ are known
at small distance, one may numerically integrate the radial Schrödinger
equation with positive energy $E$ from the known $\uzero$ and $\uinfty$
at small distance, outwards to large distances. This gives the long-range
oscillations $\left(\QDTDklzero\right)^{-1}\sin(kr+\QDTetazero-\ell\pi/2)$
and $\left(\QDTZk\right)^{-1}\sin(kr+\QDTetainf-\ell\pi/2)$, from
which $\QDTDklzero$, $\QDTetazero$, $\QDTZk$, and $\QDTetainf$
can be extracted. 

To calculate $\QDTlambdakappa$, one can start at large distance from
the exponentially decaying form $\exp(-\kappa r)$ and integrate inwards
with negative energy $E=-\hbar^{2}\kappa^{2}/2\mu$ to find the short-distance
oscillations $\uzero-\QDTlambdakappa\uinfty$ and extract $\QDTlambdakappa$.
Alternatively, one can calculate the bound state spectrum of the potential
$V$ in the $\ell$th wave for different values of the scattering
length $\QDTa$ set by altering the short-range part of $V$. In all
cases, the universal functions $\QDTetazero,\QDTetainf,\QDTDklzero,\QDTZk,$
and $\QDTlambdakappa$ can be  easily obtained with these numerical
procedures.

\begin{figure*}
\hfill{}\includegraphics[height=9cm]{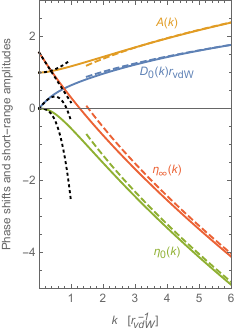}\hfill{}\includegraphics[height=9cm]{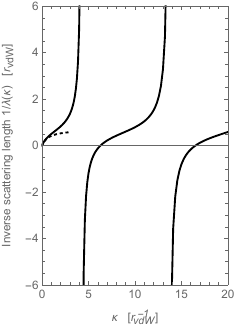}\hfill{}

\caption{\label{fig:Universal-functions}Universal functions in the s wave
for potentials with van der Waals tail $V_{\text{tail}}(r)=-C_{6}/r^{6}$.
Left panel: functions $\QDTetazero(k),\QDTetainf(k),\QDTDklzero(k)$,
and $\QDTZk(k)$ for positive energies $\frac{\hbar^{2}k^{2}}{2\mu}$.
Right panel: function $1/\QDTlambdakappa$ for negative energies $-\frac{\hbar^{2}\kappa^{2}}{2\mu}$.
The quantity $\QDTlambdakappa$ is simply the scattering length for
which the potential admits a bound state with binding energy $\frac{\hbar^{2}\kappa^{2}}{2\mu}$.
All quantities are plotted in units of the van der Waals length $r_{\text{vdW}}=\frac{1}{2}\left(2\mu C_{6}/\hbar^{2}\right)^{1/4}$.
The dotted curves correspond to the small-$k$ formulas Eqs.~(\ref{eq:etak0-low-energy}-\ref{eq:Zk-low-energy})
and Eq.~(\ref{eq:lambdak-low-energy}) and the dashed curves correspond
the large-$k$ formulas Eqs.~(\ref{eq:etak0-high-energy}-\ref{eq:Zk-high-energy}).}
\end{figure*}

\subsubsection{Case of van der Waals interactions\label{subsec:Case-of-van-der-Waals}}

For potentials with a van der Waals tail $-C_{6}/r^{6}$, the characteristic
length scale is the van der Waals length $r_{\text{vdW}}=\frac{1}{2}\left(2\mu C_{6}/\hbar^{2}\right)^{1/4}$
or equivalently the mean scattering length $\bar{a}=\frac{4\pi}{\Gamma(1/4)^{2}}r_{\text{vdW}}$.
One can in principle obtain analytical expressions of the universal
functions from the analytical solution of the Schrödinger equation
for van der Waals potentials~\cite{Gao1998}, although they are rather
involved. Alternatively, one can employ the numerical method sketched
above. Figure~\ref{fig:Universal-functions} shows the result for
the s wave. 

The functions for the s wave admit the following analytical expressions
for small $k\ll\bar{a}^{-1}$:
\begin{align}
\QDTetazero(k) & =-\frac{8}{3}\bar{a}r_{\text{vdW}}k^{3}+O(k^{4})\label{eq:etak0-low-energy}\\
\QDTetainf(k) & =\frac{\pi}{2}-\frac{4}{3}\frac{r_{\text{vdW}}^{2}}{\bar{a}}k+O(k^{3})\label{eq:etak-infinity-low-energy}\\
\QDTDklzero(k) & =k-\frac{4}{3}r_{\text{vdW}}^{2}k^{3}+O(k^{4})\label{eq:Dk0-low-energy}\\
\QDTZk(k) & =1+\frac{4}{3}\left(1-\frac{2}{3}\frac{r_{\text{vdW}}^{2}}{\bar{a}^{2}}\right)k^{2}r_{\text{vdW}}^{2}+O(k^{3})\label{eq:Zk-low-energy}
\end{align}

From this and using Eq.~(\ref{eq:tandeltak-expression}), one can
perform the effective range expansion:

\begin{align}
k\cot\QDTetakla & =-\frac{1}{\QDTa}+\frac{1}{2}r_{\text{eff}}k^{2}+O\left(k^{3}\right)\label{eq:deltak-low-energy}
\end{align}
yielding the effective range
\begin{equation}
r_{\text{eff}}=r_{\text{eff}}^{(\infty)}\left[\left(\frac{\bar{a}}{\QDTa}\right)^{2}+\left(\frac{\bar{a}}{\QDTa}-1\right)^{2}\right]\label{eq:effective-range-single-channel-vdW}
\end{equation}
with $r_{\text{eff}}^{(\infty)}=\frac{8}{3}\frac{r_{\text{vdW}}^{2}}{\bar{a}}$.

For negative energies, one finds for small $\kappa\ll\bar{a}^{-1}$:
\begin{equation}
\lambda(\kappa)=\frac{1}{\kappa}+\frac{1}{2}r_{\text{eff}}^{(\infty)}+O(\kappa)\label{eq:lambdak-low-energy}
\end{equation}

One can also derive the following expressions in the high-energy limit
$k\gg\bar{a}^{-1}$ using the WKB approximation:
\begin{align}
\QDTetazero(k) & =-\xi\times\left(kr_{\text{vdW}}\right)^{2/3}+5\pi/8\label{eq:etak0-high-energy}\\
\QDTetainf(k) & =-\xi\times\left(kr_{\text{vdW}}\right)^{2/3}+7\pi/8\label{eq:etak-infinity-high-energy}\\
\QDTDklzero(k) & =\sqrt{\frac{k}{2\bar{a}}}\label{eq:Dk0-high-energy}\\
\QDTZk(k) & =\sqrt{k\bar{a}}\label{eq:Zk-high-energy}
\end{align}
with $\xi=-2^{-1/3}\pi^{-1/2}\Gamma\left(-\frac{1}{3}\right)\Gamma\left(\frac{5}{6}\right)\approx2.0533$.

\subsubsection{Case of contact interactions\label{subsec:Case-of-contact}}

The case of contact interactions can be obtained by taking the limit
$r_{\text{vdW}}\to0$ of Eqs.~(\ref{eq:etak0-low-energy}-\ref{eq:Zk-low-energy})
and (\ref{eq:lambdak-low-energy}), yielding:
\begin{align}
\QDTetazero(k) & =0\label{eq:etak0-low-energy-1}\\
\QDTetainf(k) & =\frac{\pi}{2}\label{eq:etak-infinity-low-energy-1}\\
\QDTDklzero(k) & =k\label{eq:Dk0-low-energy-1}\\
\QDTZk(k) & =1\label{eq:Zk-low-energy-1}\\
\QDTlambdakappa & =1/\kappa\label{eq:lambdak-low-energy1}
\end{align}

\subsection{Connection with other QDT notations\label{subsec:Connection-with-other-notations}}

In the works of Refs.~\cite{Gao2005,Gao2009,Gao2011}, the quantum
defect theory is formulated with a set of four functions $Z_{ff},Z_{gg},Z_{fg},$
and $Z_{gf}$ along with a short-distance $K_{0}^{0}$ that is related
to the scattering length $a$, such that the scattering phase shift
reads:
\begin{align}
\tan\QDTetakla & =\frac{K_{0}^{0}Z_{gg}-Z_{fg}}{Z_{ff}-K_{0}^{0}Z_{gf}}\,\qquad\text{with }K_{0}^{0}=\left(1-\frac{a}{\bar{a}}\right)^{-1}\label{eq:PhaseShiftBoGao}
\end{align}
Therefore, the functions $\QDTetazero$, $\QDTetainf$, $\QDTDklzero$,
and $\QDTZk$ are related to these functions by the relations:
\begin{align}
\QDTetazero & =\arctan\frac{Z_{gg}-Z_{fg}}{Z_{ff}-Z_{gf}}\label{eq:relation1}\\
\QDTetainf & =-\arctan\frac{Z_{fg}}{Z_{ff}}\label{eq:relation2}\\
\left(\QDTDklzero\right)^{-1} & =\bar{a}\sqrt{\left(Z_{gg}-Z_{fg}\right)^{2}+\left(Z_{ff}-Z_{gf}\right)^{2}}\label{eq:relation3}\\
\left(\QDTZk\right)^{-1} & =\sqrt{Z_{ff}^{2}+Z_{fg}^{2}}\label{eq:relation4}
\end{align}
and conversely,
\begin{align}
Z_{gg} & =\bar{a}^{-1}\left(\QDTDklzero\right)^{-1}\sin\QDTetazero-\left(\QDTZk\right)^{-1}\sin\QDTetainf\label{eq:inverseRelation1}\\
Z_{fg} & =-\left(\QDTZk\right)^{-1}\sin\QDTetainf\label{eq:inverseRelation2}\\
Z_{gf} & =\left(\QDTZk\right)^{-1}\cos\QDTetainf-\bar{a}^{-1}\left(\QDTDklzero\right)^{-1}\cos\QDTetazero\label{eq:inverseRelation3}\\
Z_{ff} & =\left(\QDTZk\right)^{-1}\cos\QDTetainf\label{eq:inverseRelation4}
\end{align}

Note that the four functions $\QDTetazero$, $\QDTetainf$, $\QDTDklzero$,
and $\QDTZk$ shown in Fig.~\ref{fig:Universal-functions} all have
a simple monotonic variation with $k$, whereas the four functions
$Z_{ff},Z_{gg},Z_{fg},$ and $Z_{gf}$ have oscillatory variations.

In Refs~\cite{Ruzic2013,Julienne2006,Naidon2019}, the short-distance
energy-independent radial functions were connected to the long-range
energy-normalised radial functions through a phase shift $\varphi$
and two amplitudes $A_{k}^{-1/2}$ and $\mathcal{G}_{k},$ also denoted
as $C(E)$ and $\tan\lambda(E)$. With the current notations, $\varphi$
is the quantum defect related to $a$ through:
\begin{equation}
\tan\varphi=K_{0}^{0}=\left(1-\frac{a}{\bar{a}}\right)^{-1}\label{eq:lambda}
\end{equation}
and $C_{k}$ and $\mathcal{G}_{k}$ are related to $\QDTDkla$, $\QDTPkla$,
and $\QDTbkla$ by the relations:

\begin{align}
\QDTDkla(k) & =\left(C_{k}\right)^{-1}\sqrt{\frac{k/\bar{a}}{1+(1-a/\bar{a})^{2}}}\label{eq:Dkla-in-terms-of-CG}\\
\QDTPkla(k) & =-\sqrt{\frac{k/\bar{a}}{1+(1-r_{0})^{2}}}\left(\mathcal{G}_{k}+1-\frac{a}{\bar{a}}\right)C_{k}\label{eq:Pkla-in-terms-of-CG}\\
\QDTbkla(k) & =\bar{a}\frac{\frac{a}{\bar{a}}(\mathcal{G}_{k}-1)+2}{\mathcal{G}_{k}+1-\frac{a}{\bar{a}}}\label{eq:bkla-in-terms-of-CG}
\end{align}

\section{Renormalised Quantum Defect Theory of the isolated resonance\label{sec:Renormalised-QDT}}

\subsection{Width and shift\label{subsec:Width-and-shift}}

Combining the results of the two preceding sections, one can now formulate
the quantum defect theory of the isolated resonance. According to
Eqs.~(\ref{eq:DeltaPlus}), (\ref{eq:SinglePartialWaveApproximation})
and (\ref{eq:partial-wave-Greens-function}), the complex shift $\Delta^{+}$
is given by 
\begin{equation}
\Delta^{+}=-\frac{2\mu}{\hbar^{2}k}\int_{0}^{\infty}dr\int_{0}^{\infty}dr^{\prime}w(r)\QDTuok(r_{<})\QDTvokplus(r_{>})w(r^{\prime})\label{eq:Deltaplus1}
\end{equation}
Now, assuming that the coupling $w(r)$ is localised in the region
$r_{0}\lesssim r\ll r_{\text{tail}}$ where Eqs.~(\ref{eq:Expansionuk}-\ref{eq:Expansionvkplus})
can be used, and using Eq.~(\ref{eq:RelationBetween_Dk_bkplus_a_Pkplus})
one finds
\begin{equation}
\Delta^{+}=\frac{2\mu}{\hbar^{2}}\frac{\left(\myX-a_{\text{o}}\myY\right)\left(\myX-\QDTbklplus\myY\right)-(\QDTbklplus-a_{\text{o}})\myZ}{\QDTbklplus-a_{\text{o}}}\label{eq:Deltaplus2}
\end{equation}
with
\begin{align}
\myX & \equiv\int_{0}^{\infty}dr\;w(r)\uzero(r)\label{eq:A}\\
\myY & \equiv\int_{0}^{\infty}dr\;w(r)\uinfty(r)\label{eq:B}\\
\myZ & \equiv\int_{0}^{\infty}dr\int_{r}^{\infty}dr^{\prime}w(r)w(r^{\prime})\times\label{eq:C}\\
 & \qquad\Big(\uzero(r)\uinfty(r^{\prime})-\uinfty(r)\uzero(r^{\prime})\Big)\nonumber 
\end{align}
Thus, introducing the lengths 
\begin{align}
a_{\text{c}} & \equiv\myX/\myY,\label{eq:ac-definition}\\
a_{\text{c}}^{\prime} & \equiv\myZ/\myY^{2},\label{eq:atildec-definition}
\end{align}
 one obtains 
\begin{equation}
\boxed{\Delta^{+}=\frac{\QDTalpha}{\QDTbklplus(k)-a_{\text{o}}}+\Delta_{0}}\label{eq:Deltaplus-expression}
\end{equation}
with 
\begin{align}
\Delta_{0} & \equiv\frac{2\mu}{\hbar^{2}}\myY^{2}\left(a_{\text{o}}-a_{\text{c}}-a_{\text{c}}^{\prime}\right)\label{eq:Delta0-expression}\\
\gamma & \equiv\frac{2\mu}{\hbar^{2}}\myY^{2}\left|a_{\text{c}}-\QDTa_{\text{o}}\right|^{2}=\frac{2\mu}{4\pi\hbar^{2}}W_{0}^{2}\left(1-\frac{a_{\text{o}}}{a_{\text{c}}}\right)\label{eq:gamma-expression}
\end{align}
where the quantity $W_{0}\equiv\sqrt{4\pi}(w|\uzero)$ characterises
the strength of the coupling between the open and closed channels.
Note that $\Delta_{0}=\lim_{k\to0}\Delta$ in the case of the s wave
($\ell=0$), for which $\QDTbklplus(k)\xrightarrow[k\to0]{}\infty$.

The simplicity of Eq.~(\ref{eq:Deltaplus-expression}) is striking,
as the dependence on the closed-channel parameters $W_{0}$, $a_{\text{c}}$,
and $a_{\text{c}}^{\prime}$ is entirely encapsulated in $\Delta_{0}$
and $\QDTalpha$, while the dependence on the open-channel parameters
only appears through the scattering length $a_{\text{o}}$ in the
denominator of Eq.~(\ref{eq:Deltaplus-expression}). For energies
$E$ below the open-channel threshold $E_{\text{o}}$, the shift $\Delta^{+}$
and the length $b^{+}(ik)=\QDTlambdakappa$ are real, leading to the
simple result:
\begin{equation}
\Delta=\frac{\QDTalpha}{\QDTlambdakappa-a_{\text{o}}}+\Delta_{0}\label{eq:Delta-below-threshold}
\end{equation}

For energies $E$ above the open-channel threshold $E_{\text{o}}$,
the real and imaginary parts of $\Delta^{+}=\Delta-i\Gamma/2$ can
be obtained from Eq.~(\ref{eq:Deltaplus-expression}) using Eq.~(\ref{eq:useful-relations}):
\begin{align}
\Delta & =\frac{\QDTalpha}{\QDTbklo-a_{\text{o}}}+\Delta_{0}=-\QDTalpha\frac{\QDTDklo\QDTPklo}{k}+\Delta_{0}\label{eq:Delta-above-threshold}\\
\frac{\Gamma}{2} & =\gamma\frac{\left(\QDTDklo\right)^{2}}{k}\label{eq:Gamma-above-threshold}
\end{align}
Using the expressions Eqs.~(\ref{eq:Dkla-in-terms-of-Z-eta}-\ref{eq:Pkla-in-terms-of-Z-eta})
one finds
\begin{align}
\Delta & =\Delta_{0}+\frac{\Gamma}{2}\QDTX\label{eq:Delta-above-threshold2}\\
\frac{\Gamma}{2} & =\gamma\frac{k}{\left(\QDTZk\right)^{2}\left[1+\left(\QDTX\right)^{2}\right]}\label{eq:Gamma-above-threshold2}
\end{align}

\subsection{Low energy\label{subsec:Low-energy}}

In the low-energy limit, the general effective-range expansion of
a resonance is given by Eq.~(\ref{eq:general-effective-range}).

In the case of a resonance with van der Waals interaction, it can
be found from Eqs.~(\ref{eq:etak0-low-energy}-\ref{eq:Zk-low-energy})
that
\begin{align}
\alpha & =\gamma\left(\frac{1}{2}r_{\text{eff}}^{(\infty)}-a_{\text{o}}\right),\label{eq:alpha-for-vdW}\\
\beta & =r_{\text{eff}}^{(\infty)}\left(a_{\text{o}}-\bar{a}\right)-a_{\text{o}}^{2}.\label{eq:beta-for-vdW}
\end{align}
In the isolated resonance limit where non-resonant contributions are
negligible (i.e. $\tilde{\gamma}=\gamma$, $\tilde{\beta}=\beta$,
and $a_{\text{o}}=a_{\text{bg}}$), this leads to the effective range,
\begin{align}
r_{\text{eff}} & =\left(r_{\text{eff}}^{(\infty)}-2R_{\star}\right)\left(1-\frac{a_{\text{bg}}}{a}\right)^{2}+r_{\text{bg}}\frac{a_{\text{bg}}^{2}}{a^{2}}\label{eq:effective-range-for-vdW}\\
 & \quad+2r_{\text{eff}}^{(\infty)}\frac{a_{\text{bg}}}{a}\left(1-\frac{\bar{a}}{a_{\text{bg}}}\right)\left(1-\frac{a_{\text{bg}}}{a}\right)\nonumber 
\end{align}
where $r_{\text{eff}}^{(\infty)}=\frac{8}{3}\frac{r_{\text{vdW}}^{2}}{\bar{a}}$
and $r_{\text{bg}}$ is the open-channel effective range given in
terms of $a_{\text{bg}}$ by Eq.~(\ref{eq:effective-range-single-channel-vdW}).
It follows that close to the resonance ($a\to\infty$), the effective
range reduces to:
\begin{equation}
\boxed{r_{\text{eff}}=r_{\text{eff}}^{(\infty)}-2R_{\star}}\label{eq:effective-range-for-vdW-at-unitarity}
\end{equation}
One can see from this formula that there are two opposite limits:
when $R_{\star}\ll r_{\text{eff}}^{(\infty)}\sim r_{\text{vdW}}$
(open-channel dominated resonance, a.k.a ``broad'' resonance~\cite{Chin2010})
the effective range $r_{\text{eff}}\approx r_{\text{eff}}^{(\infty)}$
is positive and approaches the effective range of the single-channel
van der Waals potential at unitarity (see Eq.~(\ref{eq:effective-range-single-channel-vdW})),
whereas when $R_{\star}\gg r_{\text{eff}}^{(\infty)}\sim r_{\text{vdW}}$
(closed-channel dominated resonance, a.k.a ``narrow'' resonance~\cite{Chin2010}),
the effective-range is $r_{\text{eff}}\approx-2R_{\star}$ is negative
and approaches the effective range of the zero-range two-channel model
at unitarity~\cite{Petrov2004}.

Indeed, in the case of a resonance with contact interactions, it can
be found from Eqs.~(\ref{eq:etak0-low-energy-1}-\ref{eq:Zk-low-energy-1}),
or simply by taking the limit $r_{\text{vdW}}\to0$ in Eq.~(\ref{eq:effective-range-for-vdW}),
that
\begin{equation}
r_{\text{eff}}=-2R_{\star}\left(1-\frac{a_{\text{bg}}}{a}\right)^{2},\label{eq:effective-range-for-contact}
\end{equation}
which shows that the contact (zero-range) two-channel model has a
negative effective range and thus always describes a closed-channel
dominated Feshbach resonance.

\subsection{Short-distance amplitudes\label{subsec:Short-distance-amplitudes}}

The QDT gives a simple account of the wave function inside the tail
region. The radial wave function in the open-channel component is
given by the isolated resonance theory equation~(\ref{eq:RadialEquation2}).
Assuming that the coupling $w(r)$ is localised around a distance
$r_{\text{w}}$, one can use Eqs.~(\ref{eq:non-retarded-partial-wave-Green-s-function})
and (\ref{eq:phase-shift0}) to obtain the radial wave function for
$r\gg r_{\text{w}}$:
\begin{equation}
\uol(r)\underset{r\gg r_{\text{w}}}{=}\QDTuok(r)+\tan\left(\etak-\etaok\right)\QDTvok(r)\label{eq:radial-wf-large-distance}
\end{equation}
This shows that for distances beyond the coupling region, the wave
function is proportional to the solution of the open-channel potential
with a short-distance boundary condition yielding the modified scattering
phase shift $\etak$ instead of the original phase shift $\etaok$.

One can also use Eqs.~(\ref{eq:RadialEquation2}) and (\ref{eq:non-retarded-partial-wave-Green-s-function})
to obtain the radial wave function for $r\ll r_{\text{w}}$:
\begin{equation}
\uol(r)\underset{r\ll r_{\text{w}}}{=}\left[1-\left(\zeta_{\text{nr}}+\frac{\tilde{\Gamma}/2}{E-E_{\text{b}}-\Delta}\frac{(\QDTvok\vert w)}{(\QDTuok\vert w)}\right)\right]\times\QDTuok(r)\label{eq:radial-wf-short-distance}
\end{equation}
with
\begin{equation}
\zeta_{\text{nr}}\equiv\frac{2\mu}{\hbar^{2}k}(\QDTvok\vert w_{\text{nr}})\label{eq:zeta-nr}
\end{equation}
This shows that for distances beneath the coupling region the wave
function is proportional to the unperturbed solution $\QDTuok$ of
the open-channel potential $V_{\text{o}}$. 

Now, assuming that the coupling region $r\sim r_{\text{w}}$ lies
in the range $r_{0}\lesssim r\ll r_{\text{tail}}$ where the wave
functions $\QDTuok$ and $\QDTvok$ are energy independent, one can
use the QDT formalism, namely Eqs.~(\ref{eq:Expansionuk}-\ref{eq:Expansionvk}),
to further specify the form of the radial wave function $\uol$.
For $r\gg r_{\text{w}}$, one finds that $\uol$ is proportional to
the zero-energy solution with an energy-dependent scattering length
$a_{\text{eff}}$:
\begin{equation}
\boxed{\uol(r)\underset{r_{\text{w}}\ll r\ll r_{\text{tail}}}{=}\mathcal{A}_{>}(k)\times\left(\uzero(r)-a_{\text{eff}}(k)\uinfty(r)\right)}\label{eq:radial-wf-large-distance-QDT}
\end{equation}
with the amplitude $\mathcal{A}_{>}$ and scattering length $a_{\text{eff}}$
given by:
\begin{align}
\mathcal{A}_{>}(k) & \equiv\QDTDklo+\tan\left(\etak-\etaok\right)\QDTPklo\label{eq:A_sup}\\
a_{\text{eff}}(k) & \equiv\QDTa_{\text{o}}-\frac{\tan\left(\etak-\etaok\right)\frac{k}{\QDTDklo^{2}}}{1+\tan\left(\etak-\etaok\right)\frac{\QDTPklo}{\QDTDklo}}\label{eq:a_eff}
\end{align}

For $r\ll r_{\text{w}}$, the wave function is proportional to the
zero-energy solution with the unperturbed scattering length $\QDTa_{\text{o}}$:
\begin{equation}
\boxed{\uol(r)\underset{r_{0}\lesssim r\ll r_{\text{w}}}{=}\mathcal{A}_{<}(k)\times\left(\uzero(r)-\QDTa_{\text{o}}\uinfty(r)\right)}\label{eq:radial-wf-short-distance-QDT}
\end{equation}
with the amplitude $\mathcal{A}_{<}$ given by:
\begin{align}
\mathcal{A}_{<}(k) & \equiv\QDTDklo\!-\!\QDTPklo\!\left(\xi_{\text{nr}}\frac{\QDTbklo-a_{\text{c}}^{\text{nr}}}{\QDTa_{\text{o}}-a_{c}^{\text{nr}}}+\frac{\tilde{\Gamma}/2}{E-E_{\text{b}}-\Delta}\frac{\QDTbklo-a_{\text{c}}}{\QDTa_{\text{o}}-a_{c}}\right)\label{eq:A_inf}
\end{align}
where the non-resonant closed-channel scattering length $a_{\text{c}}^{\text{nr}}$
is defined by:
\[
a_{\text{c}}^{\text{nr}}\equiv\frac{(\uzero\vert w_{\text{nr}})}{(\uinfty\vert w_{\text{nr}})}.
\]

In the s wave, for small $k$, one finds:
\begin{align}
\mathcal{A}_{>}(k) & \xrightarrow[k\to0]{}k\label{eq:A_sup-small-k}\\
\mathcal{A}_{<}(k) & \xrightarrow[k\to0]{}\mathcal{A}_{>}(k)\frac{a-a_{c}-a_{\text{nr}}\frac{a_{c}-a_{c}^{\text{nr}}}{\QDTa_{\text{o}}-a_{c}^{\text{nr}}}}{\QDTa_{\text{o}}-a_{c}}\label{eq:A_inf-small-k}\\
a_{\text{eff}}(k) & \xrightarrow[k\to0]{}a\label{eq:a_eff-small-k}
\end{align}

In the fully isolated resonance limit where the non-resonant parts
are negligible, the expressions of Eqs.~(\ref{eq:A_sup},~\ref{eq:a_eff},~\ref{eq:A_inf})
simplify to:
\begin{align}
\mathcal{A}_{>}(k) & =\QDTDklo(k)\frac{\QDTa_{\text{o}}-\QDTbklo(k)}{a_{\text{eff}}(k)-\QDTbklo(k)}\label{eq:A_sup-fully-isolated}\\
\mathcal{A}_{<}(k) & =\mathcal{A}_{>}(k)\frac{a_{\text{eff}}(k)-a_{c}}{\QDTa_{\text{o}}-a_{c}}\label{eq:A_inf-fully-isolated}\\
a_{\text{eff}}(k) & =a_{\text{o}}+\frac{\QDTalpha}{E-E_{\text{b}}-\Delta_{0}}\label{eq:a_eff-fully-isolated}
\end{align}
In particular, at low energy such that $a_{\text{eff}}(k)\approx a$,
one finds the simple formula for the ratio:
\begin{equation}
\boxed{\frac{\mathcal{A}_{<}}{\mathcal{A}_{>}}=\frac{a-a_{\text{c}}}{a_{\text{o}}-a_{\text{c}}}}\label{eq:ratio-formula}
\end{equation}
showing that the short-distance amplitude vanishes when $a=a_{\text{c}}$.
Note that the general formula Eq.~(\ref{eq:A_sup-small-k}) for a
partially isolated resonance reduces to Eq.~(\ref{eq:ratio-formula})
in the special case where $a_{\text{c}}^{\text{nr}}=a_{\text{c}}$.
This happens when the wave functions of the resonant and non-resonant
bare states in the closed channel are also energy-independent in the
coupling region, and thus characterised by the same scattering length. 

\begin{figure}[t]
\includegraphics[width=8cm]{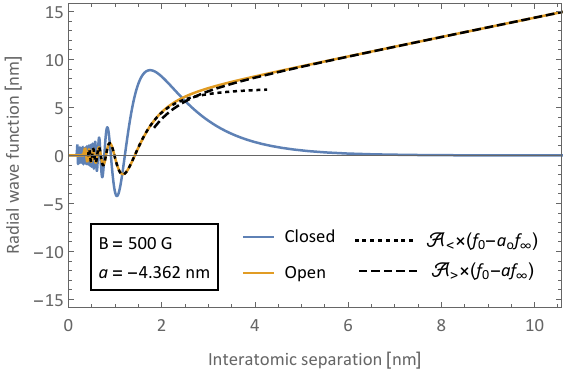}

\includegraphics[width=8cm]{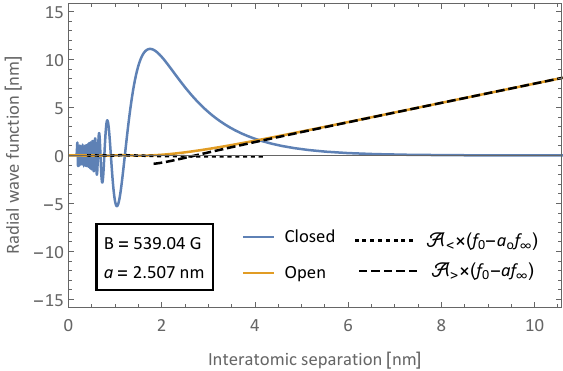}

\caption{\label{fig:Lithium-6-radial-wave-functions}Effective two-channel
radial wave functions $u_{\text{c}}$ (\textquotedblleft closed\textquotedblright ,
blue curve) and $u_{\text{o}}$ (\textquotedblleft open\textquotedblright ,
orange curve) of the lithium-6 diatomic $ab$ resonance near $B=834\ $G.
Top: wave functions at $B=500\,$G, corresponding to a scattering
length $a=-4.362$~nm. Bottom: wave functions at $B=539.04\,$G,
corresponding to a scattering length $a=a_{\text{c}}=2.507$~nm.
The open-channel wave function $u_{\text{o}}$ is fitted at large
distance by the wave function of Eq.~(\ref{eq:radial-wf-large-distance-QDT})
(dashed curve) and at short by the wave function of Eq.~(\ref{eq:radial-wf-short-distance-QDT})
(dotted curve).}
\end{figure}

\subsection{Application to lithium-6\label{subsec:Application-to-lithium-6}}

\begin{figure*}
\includegraphics[height=6cm]{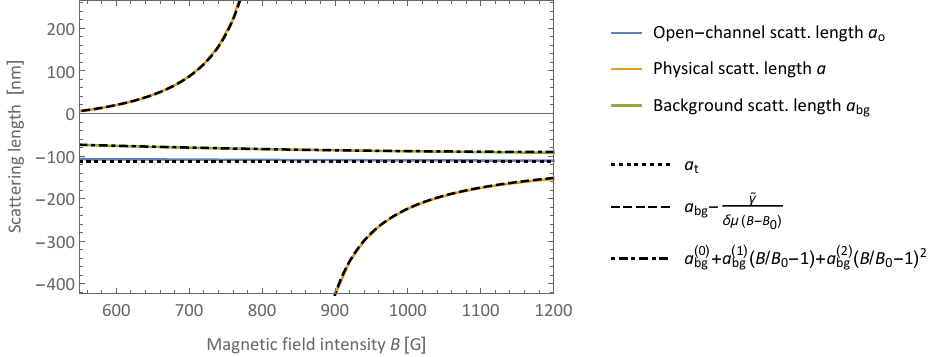}\caption{\label{fig:lithium6-ScatteringLengths}Open-channel scattering length
$a_{\text{o}}$ (orange curve), physical scattering length $a$ (blue
curve), and background scattering length $a_{\text{bg}}$ of the lithium-6
diatomic $ab$ resonance near $B=834\ $G, obtained by fitting the
numerical wave functions, as a function of magnetic field intensity.
The open-channel scattering length $a_{\text{o}}$ is close to the
triplet scattering length $a_{\text{t}}=-112.8\,$nm (dotted line),
while the physical scattering length is well reproduced by Eq.~(\ref{eq:PhysicalScatteringLength})
(dahsed curve) with $\tilde{\gamma}/h=62\,770\,$MHz~nm, and $\tilde{E}_{b}=E_{b}+\Delta_{0}-E_{\text{o}}=\delta\mu\times(B-B_{0})$
with $\delta\mu/h=2.8\,$MHz/G and $B_{0}=834.08$~G. This yields
the background scattering length $a_{\text{bg}}$, which is well reproduced
by Eq.~(\ref{eq:TaylorExpansionOfa_bg}).}
\end{figure*}
The lithium-6 $ab$ diatomic resonance (where $ab$ designates the
two lowest hyperfine states of lithium-6) near the magnetic field
intensity $B=834\,$G is described by five hyperfine channels characterised
by a total spin projection $m_{F}=0$. The interaction between the
atoms depends on the total electronic spin $S$ of the two valence
electrons, which can be either in a singlet ($S=0$) or triplet ($S=1$)
state. This multi-channel system with radial components $u_{i}(r)$
($i=1,\dots,5$) can thus be solved numerically using the singlet
and triplet interaction potentials and the atomic hyperfine Hamiltonian. 

The bare bound state causing this resonance has been identified as
the $\nu=38$ s-wave level of the singlet interaction potential, with
radial wave function $u_{\text{b}}(r)$. Therefore, to construct the
effective two-channel components, one can project the components $u_{i}$
onto the bare bound state to obtain the closed-channel component $u_{\text{c}}$,
and project out the bare bound state and retain only the $ab$ entrance
component ($i=1$) to obtain the open-channel component $u_{\text{o}}$.
Explicitly,
\begin{align}
u_{\text{c}}(r) & =\sqrt{\sum_{i,j=1}^{5}\left|\alpha_{i,j}(u_{\text{b}}\vert u_{j})\right|^{2}}u_{\text{b}}(r)\label{eq:uc-effective}\\
u_{\text{o}}(r) & =u_{1}(r)-\sum_{j=1}^{5}\alpha_{1j}(u_{\text{b}}\vert u_{j})u_{\text{b}}(r)\label{eq:uo-effective}
\end{align}
where $\alpha_{ij}$ are the matrix elements of the projector $1-\hat{S}^{2}$
onto the singlet state.

The zero-energy components are shown in Fig.~\ref{fig:Lithium-6-radial-wave-functions}
for two different values of the magnetic field intensity. The open-channel
wave function $u_{\text{o}}$ (orange curve) is well fitted at large
distance by the wave function of Eq.~(\ref{eq:radial-wf-large-distance-QDT})
(dashed curve), and at short distance by the wave function of Eq.~(\ref{eq:radial-wf-short-distance-QDT})
(dotted curve). The two fits deviate from $u_{\text{o}}$ in a region
of distances around $r_{\text{w}}=2.6$~nm, which shows that the
inter-channel coupling is localised in that region. The QDT is therefore
an accurate description for energies smaller than 240~$h$MHz$\,\sim$~10~mK
above and below the threshold. The fits enable to extract the amplitudes
$\mathcal{A}_{>}$ and $\mathcal{A}_{<}$, as well as the open-channel
scattering length $a_{\text{o}}$ and the physical scattering length
$a$.

Both scattering lengths $a_{\text{o}}$ and $a$ are plotted as a
function of magnetic field intensity as blue and orange curves in
Fig.~\ref{fig:lithium6-ScatteringLengths}. One can see that the
open-channel scattering length $a_{\text{o}}$ is close to the triplet
scattering length $a_{\text{t}}=-112.8\,$nm, confirming the spin
triplet character of the open channel, while the physical scattering
length $a$ is well reproduced by the formula of Eq.~(\ref{eq:PhysicalScatteringLength}).
This yields the background scattering length $a_{\text{bg}}$, which
has a small dependence on the magnetic field as shown by the green
curve in Fig.~\ref{fig:lithium6-ScatteringLengths}. This dependence
is captured by the following Taylor expansion around $B_{0}$:
\begin{equation}
a_{\text{bg}}=a_{\text{bg}}^{(0)}+a_{\text{bg}}^{(1)}\left(B/B_{0}-1\right)+a_{\text{bg}}^{(2)}\left(B/B_{0}-1\right)^{2}\label{eq:TaylorExpansionOfa_bg}
\end{equation}
with $a_{\text{bg}}^{(0)}=-84.89$~nm, $a_{\text{bg}}^{(1)}=-24.19$~nm,
and $a_{\text{bg}}^{(2)}=22.77$~nm.

\begin{figure}[t]
\includegraphics[viewport=0bp 0bp 300bp 198bp,width=8cm]{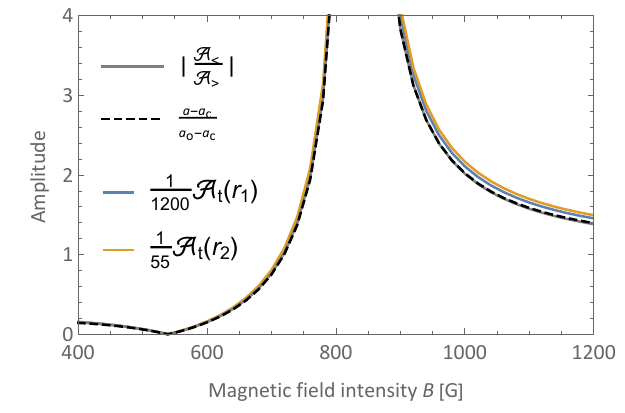}

\includegraphics[width=8cm]{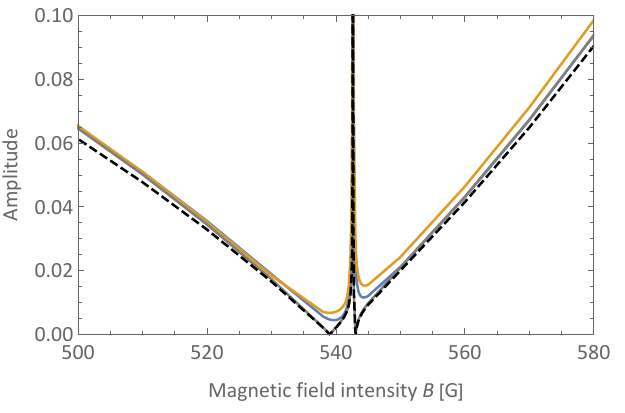}

\caption{\label{fig:lithium6-amplitudes}Top: Short-distance amplitude of the
lithium-6 diatomic open channel as a function of magnetic field intensity.
The grey curve shows the ratio of the amplitudes $\mathcal{A}_{<}$
and $\mathcal{A}_{>}$ of Eqs.~(\ref{eq:radial-wf-large-distance-QDT}-\ref{eq:radial-wf-short-distance-QDT}).
It is well reproduced by Eq.~(\ref{eq:ratio-formula}) with $a_{\text{c}}=2.510$~nm
(dashed curve). The blue and orange curves show the triplet amplitude
$\mathcal{A}_{\text{t}}(r)$ of Eq.~(\ref{eq:triplet-amplitude})
at the probing distance $r_{1}=1.9$~nm (blue) and $r_{2}=0.8$~nm
(orange). Bottom: closeup of the top figure in the region where the
short-distance amplitude vanishes.}

\end{figure}

The ratio $\vert\mathcal{A}_{<}/\mathcal{A}_{>}\vert$ is plotted
in Fig.~\ref{fig:lithium6-amplitudes} as a function of magnetic
field intensity. It is well reproduced by the formula Eq.~(\ref{eq:ratio-formula})
with a closed-channel scattering length $a_{\text{c}}=2.507$~nm~$\approx a_{\text{s}}$
and an open-channel scattering length $a_{\text{o}}=-109.6$~nm~$\approx a_{\text{t}}$.
It should be noted that the non-resonant contribution is not negligible
for this resonance: the background scattering length $a_{\text{bg}}$
is found to be around -85~nm, significantly differing from $a_{\text{o}}$.
Thus, in principle one may not use Eq.~(\ref{eq:ratio-formula})
which is obtained in the fully isolated resonance limit, but Eq.~(\ref{eq:A_inf-small-k}),
which includes the non-resonant correction. However, it turns out
that $a_{\text{o}}$ is very large compared to $a_{\text{c}}$ and
$a_{\text{c}}^{\text{nr}}$ (both are presumably of the same order,
or even possibly equal) so that Eq.~(\ref{eq:ratio-formula}) is
a good approximation of Eq.~(\ref{eq:A_inf-small-k}) in this case.

One can see that the short-distance amplitude vanishes at the magnetic
field intensity $B=539.04\,$G corresponding as expected to $a=a_{\text{c}}$.
This suppression of the open-channel amplitude at short distance can
be visualised in the bottom panel of Fig.~\ref{fig:Lithium-6-radial-wave-functions}.
A close look around this magnetic field (see the bottom panel of Fig.~\ref{fig:lithium6-amplitudes})
reveals that a very narrow resonance accidentally occurs close to
that point. Although the presence of this extra resonance, which is
due to a bound state with total nuclear spin $I=2$~\cite{Chin2010},
complicates a bit the variation of $\vert\mathcal{A}_{<}/\mathcal{A}_{>}\vert$,
it is still reproduced by Eq.~(\ref{eq:ratio-formula}) with the
same value of $a_{\text{c}}$ when the precise variation of $a$ (inluding
the narrow resonance) is taken into account. This is because both
the broad and narrow resonances originate from the same bound state
$\nu=38$ of the singlet potential, thus having the same values of
$a_{\text{c}}$. The fact that the obtained value of $a_{\text{c}}=2.507$~nm
is very close to the singlet scattering length $a_{\text{s}}=2.391$~nm
confirms that the singlet character of the closed-channel bound state.

Since the open channel corresponds essentially to the triplet component,
an experiment probing the triplet component, for instance by photoassociation,
could reveal how the amplitude of the open-channel wave function vanishes
near $a=a_{\text{c}}$. This is illustrated in Fig.~\ref{fig:lithium6-amplitudes},
where the triplet amplitude 
\begin{equation}
\mathcal{A}_{\text{t}}(r)\equiv\sqrt{\sum_{i,j=1}^{5}\left|(1-\alpha_{i,j})u_{j}(r)\right|^{2}}\label{eq:triplet-amplitude}
\end{equation}
at two different probing distances $r$ is plotted as a function of
the magnetic field intensity. For both probing distances, the triplet
amplitude reproduces very well the short-distance amplitude $\mathcal{A}_{<}$
up to a scaling factor. However, very close to the points where $\mathcal{A}_{<}$
vanishes, the triplet amplitude does not completely vanish, an indication
that the triplet component does not perfectly account for the open
channel, but also includes non-vanishing admixtures. Nevertheless,
the measurement of $\mathcal{A}_{<}$ over a wide enough range of
magnetic field intensities would enable to determine $a_{\text{c}}$.\vspace{1cm}

\supplementalBibliography

\clearpage

\bibliographystyle{IEEEtran2}
\bibliography{paper40}

\begin{thebibliography}{10}
\providecommand{\url}[1]{#1}
\csname url@samestyle\endcsname
\providecommand{\newblock}{\relax}
\providecommand{\bibinfo}[2]{#2}
\providecommand{\BIBentrySTDinterwordspacing}{\spaceskip=0pt\relax}
\providecommand{\BIBentryALTinterwordstretchfactor}{4}
\providecommand{\BIBentryALTinterwordspacing}{\spaceskip=\fontdimen2\font plus
\BIBentryALTinterwordstretchfactor\fontdimen3\font minus \fontdimen4\font\relax}
\providecommand{\BIBforeignlanguage}[2]{{%
\expandafter\ifx\csname l@#1\endcsname\relax
\typeout{** WARNING: IEEEtran.bst: No hyphenation pattern has been}%
\typeout{** loaded for the language `#1'. Using the pattern for}%
\typeout{** the default language instead.}%
\else
\language=\csname l@#1\endcsname
\fi
#2}}
\providecommand{\BIBdecl}{\relax}
\BIBdecl

\bibitem{Feshbach1958}
\BIBentryALTinterwordspacing
H.~Feshbach, ``Unified theory of nuclear reactions''  \href{https://www.sciencedirect.com/science/article/pii/0003491658900071}{ \emph{Annals of Physics}, {\bfseries ~5},  357--390, 1958}.
\BIBentrySTDinterwordspacing

\bibitem{Fano1961}
\BIBentryALTinterwordspacing
U.~Fano, ``Effects of Configuration Interaction on Intensities and Phase Shifts''  \href{https://link.aps.org/doi/10.1103/PhysRev.124.1866}{ \emph{Phys. Rev.}, {\bfseries  124},  1866--1878, 1961}.
\BIBentrySTDinterwordspacing

\bibitem{Chin2010}
\BIBentryALTinterwordspacing
C.~Chin, R.~Grimm, P.~Julienne, and E.~Tiesinga, ``Feshbach resonances in ultracold gases''  \href{https://link.aps.org/doi/10.1103/RevModPhys.82.1225}{ \emph{Rev. Mod. Phys.}, {\bfseries ~82},  1225--1286, 2010}.
\BIBentrySTDinterwordspacing

\bibitem{Kokkelmans2014}
\BIBentryALTinterwordspacing
S.~Kokkelmans, ``Feshbach Resonances in Ultracold Gases'' in  \href{https://www.worldscientific.com/doi/abs/10.1142/9781783264766_0004}{ \emph{Quantum Gas Experiments: Exploring Many-Body States}, ser. Cold Atoms, ch.~4,  63--85.\hskip 1em plus 0.5em minus 0.4em\relax World Scientific, 2014}.

\bibitem{Tiesinga1993}
\BIBentryALTinterwordspacing
E.~Tiesinga, B.~J. Verhaar, and H.~T.~C. Stoof, ``Threshold and resonance phenomena in ultracold ground-state collisions''  \href{https://link.aps.org/doi/10.1103/PhysRevA.47.4114}{ \emph{Phys. Rev. A}, {\bfseries ~47},  4114--4122, 1993}.
\BIBentrySTDinterwordspacing

\bibitem{Inouye1998}
\BIBentryALTinterwordspacing
S.~Inouye, M.~R. Andrews, J.~Stenger, H.-J. Miesner, D.~M. Stamper-Kurn, and W.~Ketterle, ``Observation of Feshbach resonances in a Bose-Einstein condensate''  \href{http://dx.doi.org/10.1038/32354}{ \emph{Nature}, {\bfseries  392},  151--154, 1998}.
\BIBentrySTDinterwordspacing

\bibitem{Hyodo2013}
\BIBentryALTinterwordspacing
T.~Hyodo, ``Structure and compositeness of hadron resonances''  \href{https://doi.org/10.1142/S0217751X13300457}{ \emph{Int. J. Mod. Phys. A}, {\bfseries ~28},  1330045, 2013}.
\BIBentrySTDinterwordspacing

\bibitem{Dong2021}
\BIBentryALTinterwordspacing
X.-K. Dong, F.-K. Guo, and B.-S. Zou, ``Explaining the many threshold structures in the heavy-quark hadron spectrum''  \href{https://link.aps.org/doi/10.1103/PhysRevLett.126.152001}{ \emph{Phys. Rev. Lett.}, {\bfseries  126},  152001, 2021}.
\BIBentrySTDinterwordspacing

\bibitem{Braaten2005}
\BIBentryALTinterwordspacing
E.~Braaten and M.~Kusunoki, ``Factorization in the production and decay of the $X(3872)$''  \href{https://link.aps.org/doi/10.1103/PhysRevD.72.014012}{ \emph{Phys. Rev. D}, {\bfseries ~72},  014012, 2005}.
\BIBentrySTDinterwordspacing

\bibitem{Kuhlenkamp2022}
\BIBentryALTinterwordspacing
C.~Kuhlenkamp, M.~Knap, M.~Wagner, R.~Schmidt, and A.~m.~c. Imamo\ifmmode~\breve{g}\else \u{g}\fi{}lu, ``Tunable Feshbach Resonances and Their Spectral Signatures in Bilayer Semiconductors''  \href{https://link.aps.org/doi/10.1103/PhysRevLett.129.037401}{ \emph{Phys. Rev. Lett.}, {\bfseries  129},  037401, 2022}.
\BIBentrySTDinterwordspacing

\bibitem{Tajima2024}
\BIBentryALTinterwordspacing
H.~Tajima, H.~Aoki, A.~Perali, and A.~Bianconi, ``Emergent Fano-Feshbach resonance in two-band superconductors with an incipient quasiflat band: Enhanced critical temperature evading particle-hole fluctuations''  \href{https://link.aps.org/doi/10.1103/PhysRevB.109.L140504}{ \emph{Phys. Rev. B}, {\bfseries  109},  L140504, 2024}.
\BIBentrySTDinterwordspacing

\bibitem{Chin2004}
\BIBentryALTinterwordspacing
C.~Chin, V.~Vuleti\ifmmode~\acute{c}\else \'{c}\fi{}, A.~J. Kerman, S.~Chu, E.~Tiesinga, P.~J. Leo, and C.~J. Williams, ``Precision Feshbach spectroscopy of ultracold ${\mathrm{Cs}}_{2}$''  \href{https://link.aps.org/doi/10.1103/PhysRevA.70.032701}{ \emph{Phys. Rev. A}, {\bfseries ~70},  032701, 2004}.
\BIBentrySTDinterwordspacing

\bibitem{Raoult2004}
\BIBentryALTinterwordspacing
M.~Raoult and F.~H. Mies, ``Feshbach resonance in atomic binary collisions in the Wigner threshold law regime''  \href{https://link.aps.org/doi/10.1103/PhysRevA.70.012710}{ \emph{Phys. Rev. A}, {\bfseries ~70},  012710, 2004}.
\BIBentrySTDinterwordspacing

\bibitem{Schunck2005}
\BIBentryALTinterwordspacing
C.~H. Schunck, M.~W. Zwierlein, C.~A. Stan, S.~M.~F. Raupach, W.~Ketterle, A.~Simoni, E.~Tiesinga, C.~J. Williams, and P.~S. Julienne, ``Feshbach resonances in fermionic $^{6}\mathrm{Li}$''  \href{https://link.aps.org/doi/10.1103/PhysRevA.71.045601}{ \emph{Phys. Rev. A}, {\bfseries ~71},  045601, 2005}.
\BIBentrySTDinterwordspacing

\bibitem{Gao2005}
\BIBentryALTinterwordspacing
B.~Gao, E.~Tiesinga, C.~J. Williams, and P.~S. Julienne, ``Multichannel quantum-defect theory for slow atomic collisions''  \href{https://link.aps.org/doi/10.1103/PhysRevA.72.042719}{ \emph{Phys. Rev. A}, {\bfseries ~72},  042719, 2005}.
\BIBentrySTDinterwordspacing

\bibitem{Hanna2009}
\BIBentryALTinterwordspacing
T.~M. Hanna, E.~Tiesinga, and P.~S. Julienne, ``Prediction of Feshbach resonances from three input parameters''  \href{https://link.aps.org/doi/10.1103/PhysRevA.79.040701}{ \emph{Phys. Rev. A}, {\bfseries ~79},  040701(R), 2009}.
\BIBentrySTDinterwordspacing

\bibitem{Jachymski2013}
\BIBentryALTinterwordspacing
K.~Jachymski and P.~S. Julienne, ``Analytical model of overlapping Feshbach resonances''  \href{https://link.aps.org/doi/10.1103/PhysRevA.88.052701}{ \emph{Phys. Rev. A}, {\bfseries ~88},  052701, 2013}.
\BIBentrySTDinterwordspacing

\bibitem{Chilcott2021}
\BIBentryALTinterwordspacing
M.~Chilcott, R.~Thomas, and N.~Kj\ae{}rgaard, ``Experimental observation of the avoided crossing of two $S$-matrix resonance poles in an ultracold atom collider''  \href{https://link.aps.org/doi/10.1103/PhysRevResearch.3.033209}{ \emph{Phys. Rev. Res.}, {\bfseries ~3},  033209, 2021}.
\BIBentrySTDinterwordspacing

\bibitem{Timmermans1999}
\BIBentryALTinterwordspacing
E.~Timmermans, P.~Tommasini, M.~Hussein, and A.~Kerman, ``Feshbach resonances in atomic Bose-Einstein condensates''  \href{https://www.sciencedirect.com/science/article/pii/S0370157399000253}{ \emph{Physics Reports}, {\bfseries  315},  199--230, 1999}.
\BIBentrySTDinterwordspacing

\bibitem{Koehler2006}
\BIBentryALTinterwordspacing
T.~K\"ohler, K.~G\'oral, and P.~S. Julienne, ``Production of cold molecules via magnetically tunable Feshbach resonances''  \href{https://link.aps.org/doi/10.1103/RevModPhys.78.1311}{ \emph{Rev. Mod. Phys.}, {\bfseries ~78},  1311--1361, 2006}.
\BIBentrySTDinterwordspacing

\bibitem{Naidon2019}
\BIBentryALTinterwordspacing
P.~Naidon and L.~Pricoupenko, ``Width and shift of Fano-Feshbach resonances for van der Waals interactions''  \href{https://link.aps.org/doi/10.1103/PhysRevA.100.042710}{ \emph{Phys. Rev. A}, {\bfseries  100},  042710, 2019}.
\BIBentrySTDinterwordspacing

\bibitem{JonaLasinio2008}
\BIBentryALTinterwordspacing
M.~Jona-Lasinio, L.~Pricoupenko, and Y.~Castin, ``Three fully polarized fermions close to a $\mathit{p}$-wave Feshbach resonance''  \href{https://link.aps.org/doi/10.1103/PhysRevA.77.043611}{ \emph{Phys. Rev. A}, {\bfseries ~77},  043611, 2008}.
\BIBentrySTDinterwordspacing

\bibitem{Schmidt2012}
\BIBentryALTinterwordspacing
R.~Schmidt, S.~P. Rath, and W.~Zwerger, ``Efimov physics beyond universality''  \href{http://dx.doi.org/10.1140/epjb/e2012-30841-3}{ \emph{Eur. Phys. J. B}, {\bfseries ~85},  386, 2012}.
\BIBentrySTDinterwordspacing

\bibitem{Pricoupenko2013}
\BIBentryALTinterwordspacing
L.~Pricoupenko, ``Many Bosons in a Narrow Magnetic Feshbach Resonance''  \href{https://link.aps.org/doi/10.1103/PhysRevLett.110.180402}{ \emph{Phys. Rev. Lett.}, {\bfseries  110},  180402, 2013}.
\BIBentrySTDinterwordspacing

\bibitem{Cohen2004}
\BIBentryALTinterwordspacing
T.~D. Cohen, B.~A. Gelman, and U.~{van Kolck}, ``An effective field theory for coupled-channel scattering''  \href{https://www.sciencedirect.com/science/article/pii/S0370269304004332}{ \emph{Physics Letters B}, {\bfseries  588},  57--66, 2004}.
\BIBentrySTDinterwordspacing

\bibitem{Kinugawa2024}
\BIBentryALTinterwordspacing
T.~Kinugawa and T.~Hyodo, ``Compositeness of ${T}_{cc}$ and $X(3872)$ by considering decay and coupled-channels effects''  \href{https://link.aps.org/doi/10.1103/PhysRevC.109.045205}{ \emph{Phys. Rev. C}, {\bfseries  109},  045205, 2024}.
\BIBentrySTDinterwordspacing

\bibitem{Greene1982}
\BIBentryALTinterwordspacing
C.~H. Greene, A.~R.~P. Rau, and U.~Fano, ``General form of the quantum-defect theory. II''  \href{https://link.aps.org/doi/10.1103/PhysRevA.26.2441}{ \emph{Phys. Rev. A}, {\bfseries ~26},  2441--2459, 1982}.
\BIBentrySTDinterwordspacing

\bibitem{Mies1984}
\BIBentryALTinterwordspacing
F.~H. Mies, ``A multichannel quantum defect analysis of diatomic predissociation and inelastic atomic scattering''  \href{http://dx.doi.org/10.1063/1.447000}{ \emph{J. Chem. Phys.}, {\bfseries ~80},  2514, 1984}.
\BIBentrySTDinterwordspacing

\bibitem{Mies1984b}
\BIBentryALTinterwordspacing
F.~H. Mies and P.~S. Julienne, ``A multichannel quantum defect analysis of two-state couplings in diatomic molecules''  \href{http://dx.doi.org/10.1063/1.447046}{ \emph{J. Chem. Phys.}, {\bfseries ~80},  2526--2536, 1984}.
\BIBentrySTDinterwordspacing

\bibitem{Mies2000}
\BIBentryALTinterwordspacing
F.~H. Mies and M.~Raoult, ``Analysis of threshold effects in ultracold atomic collisions''  \href{https://link.aps.org/doi/10.1103/PhysRevA.62.012708}{ \emph{Phys. Rev. A}, {\bfseries ~62},  012708, 2000}.
\BIBentrySTDinterwordspacing

\bibitem{Gao2009}
\BIBentryALTinterwordspacing
B.~Gao, ``Analytic description of atomic interaction at ultracold temperatures: The case of a single channel''  \href{https://link.aps.org/doi/10.1103/PhysRevA.80.012702}{ \emph{Phys. Rev. A}, {\bfseries ~80},  012702, 2009}.
\BIBentrySTDinterwordspacing

\bibitem{Gao2011}
\BIBentryALTinterwordspacing
B.~Gao, ``Analytic description of atomic interaction at ultracold temperatures. II. Scattering around a magnetic Feshbach resonance''  \href{https://link.aps.org/doi/10.1103/PhysRevA.84.022706}{ \emph{Phys. Rev. A}, {\bfseries ~84},  022706, 2011}.
\BIBentrySTDinterwordspacing

\bibitem{Ruzic2013}
\BIBentryALTinterwordspacing
B.~P. Ruzic, C.~H. Greene, and J.~L. Bohn, ``Quantum defect theory for high-partial-wave cold collisions''  \href{https://doi.org/10.1103/PhysRevA.87.032706}{ \emph{Phys. Rev. A}, {\bfseries ~87},  032706, 2013}.
\BIBentrySTDinterwordspacing

\bibitem{Chilcott2022}
\BIBentryALTinterwordspacing
M.~Chilcott, J.~F.~E. Croft, R.~Thomas, and N.~Kj\ae{}rgaard, ``Microscopy of an ultranarrow Feshbach resonance using a laser-based atom collider: A quantum defect theory analysis''  \href{https://link.aps.org/doi/10.1103/PhysRevA.106.023303}{ \emph{Phys. Rev. A}, {\bfseries  106},  023303, 2022}.
\BIBentrySTDinterwordspacing

\bibitem{Gao1998}
\BIBentryALTinterwordspacing
B.~Gao, ``Quantum-defect theory of atomic collisions and molecular vibration spectra''  \href{http://link.aps.org/doi/10.1103/PhysRevA.58.4222}{ \emph{Phys. Rev. A}, {\bfseries ~58},  4222--4225, 1998}.
\BIBentrySTDinterwordspacing

\bibitem{Lee1954}
\BIBentryALTinterwordspacing
T.~D. Lee, ``Some Special Examples in Renormalizable Field Theory''  \href{https://link.aps.org/doi/10.1103/PhysRev.95.1329}{ \emph{Phys. Rev.}, {\bfseries ~95},  1329--1334, 1954}.
\BIBentrySTDinterwordspacing

\bibitem{Petrov2004}
\BIBentryALTinterwordspacing
D.~S. Petrov, ``Three-Boson Problem near a Narrow Feshbach Resonance''  \href{https://link.aps.org/doi/10.1103/PhysRevLett.93.143201}{ \emph{Phys. Rev. Lett.}, {\bfseries ~93},  143201, 2004}.
\BIBentrySTDinterwordspacing

\bibitem{Gogolin2008}
\BIBentryALTinterwordspacing
A.~O. Gogolin, C.~Mora, and R.~Egger, ``Analytical Solution of the Bosonic Three-Body Problem''  \href{https://link.aps.org/doi/10.1103/PhysRevLett.100.140404}{ \emph{Phys. Rev. Lett.}, {\bfseries  100},  140404, 2008}.
\BIBentrySTDinterwordspacing

\bibitem{Nishida2012}
\BIBentryALTinterwordspacing
Y.~Nishida, ``New Type of Crossover Physics in Three-Component Fermi Gases''  \href{https://link.aps.org/doi/10.1103/PhysRevLett.109.240401}{ \emph{Phys. Rev. Lett.}, {\bfseries  109},  240401, 2012}.
\BIBentrySTDinterwordspacing

\bibitem{Hammer2004}
\BIBentryALTinterwordspacing
H.-W. Hammer, ``Universality in the physics of cold atoms with large scattering length''  \href{https://www.sciencedirect.com/science/article/pii/S0375947404003410}{ \emph{Nuclear Physics A}, {\bfseries  737},  275--279, 2004}.
\BIBentrySTDinterwordspacing

\bibitem{Fabbietti2021}
\BIBentryALTinterwordspacing
L.~Fabbietti, V.~M. Sarti, and O.~V. Doce, ``Study of the Strong Interaction Among Hadrons with Correlations at the LHC''  \href{https://doi.org/10.1146/annurev-nucl-102419-034438}{ \emph{Annu. Rev. Nucl. Part. Sci.}, {\bfseries ~71},  377--402, 2021}.
\BIBentrySTDinterwordspacing

\bibitem{Song2022}
\BIBentryALTinterwordspacing
J.~Song, L.~R. Dai, and E.~Oset, ``How much is the compositeness of a bound state constrained by a and $r_{0}$ The role of the interaction range''  \href{Eur. Phys. J. A 58, 133 (2022). https://doi.org/10.1140/epja/s10050-022-00753-3}{ \emph{Eur. Phys. J. A}, {\bfseries ~58},  133, 2022}.
\BIBentrySTDinterwordspacing

\bibitem{Albaladejo2022}
\BIBentryALTinterwordspacing
M.~Albaladejo and J.~Nieves, ``Compositeness of S-wave weakly-bound states from next-to-leading order Weinberg’s relations''  \href{https://doi.org/10.1140/epjc/s10052-022-10695-1}{ \emph{Eur. Phys. J. C}, {\bfseries ~82},  724, 2022}.
\BIBentrySTDinterwordspacing

\bibitem{Kinugawa2022a}
\BIBentryALTinterwordspacing
T.~Kinugawa and T.~Hyodo, ``Structure of exotic hadrons by a weak-binding relation with finite-range correction''  \href{https://link.aps.org/doi/10.1103/PhysRevC.106.015205}{ \emph{Phys. Rev. C}, {\bfseries  106},  015205, 2022}.
\BIBentrySTDinterwordspacing

\bibitem{Weinberg1963}
\BIBentryALTinterwordspacing
S.~Weinberg, ``Elementary Particle Theory of Composite Particles''  \href{https://link.aps.org/doi/10.1103/PhysRev.130.776}{ \emph{Phys. Rev.}, {\bfseries  130},  776--783, 1963}.
\BIBentrySTDinterwordspacing

\bibitem{Choi2003}
\BIBentryALTinterwordspacing
S.-K. Choi \emph{et~al.}, ``Observation of a Narrow Charmoniumlike State in Exclusive ${B}^{\ifmmode\pm\else\textpm\fi{}}\ensuremath{\rightarrow}{K}^{\ifmmode\pm\else\textpm\fi{}}{\ensuremath{\pi}}^{+}{\ensuremath{\pi}}^{\ensuremath{-}}J/\ensuremath{\psi}$ Decays''  \href{https://link.aps.org/doi/10.1103/PhysRevLett.91.262001}{ \emph{Phys. Rev. Lett.}, {\bfseries ~91},  262001, 2003}.
\BIBentrySTDinterwordspacing

\bibitem{Aaij2020}
\BIBentryALTinterwordspacing
R.~Aaij \emph{et~al.}, ``Study of the lineshape of the ${\ensuremath{\chi}}_{c1}(3872)$ state''  \href{https://link.aps.org/doi/10.1103/PhysRevD.102.092005}{ \emph{Phys. Rev. D}, {\bfseries  102},  092005, 2020}.
\BIBentrySTDinterwordspacing

\bibitem{Partridge2005}
\BIBentryALTinterwordspacing
G.~B. Partridge, K.~E. Strecker, R.~I. Kamar, M.~W. Jack, and R.~G. Hulet, ``Molecular Probe of Pairing in the BEC-BCS Crossover''  \href{https://link.aps.org/doi/10.1103/PhysRevLett.95.020404}{ \emph{Phys. Rev. Lett.}, {\bfseries ~95},  020404, 2005}.
\BIBentrySTDinterwordspacing

\bibitem{Chen2005}
\BIBentryALTinterwordspacing
Q.~Chen and K.~Levin, ``Population of Closed-Channel Molecules in Trapped Fermi Gases with Broad Feshbach Resonances''  \href{https://link.aps.org/doi/10.1103/PhysRevLett.95.260406}{ \emph{Phys. Rev. Lett.}, {\bfseries ~95},  260406, 2005}.
\BIBentrySTDinterwordspacing

\bibitem{Zhang2009}
\BIBentryALTinterwordspacing
S.~Zhang and A.~J. Leggett, ``Universal properties of the ultracold Fermi gas''  \href{https://link.aps.org/doi/10.1103/PhysRevA.79.023601}{ \emph{Phys. Rev. A}, {\bfseries ~79},  023601, 2009}.
\BIBentrySTDinterwordspacing

\bibitem{Werner2009}
\BIBentryALTinterwordspacing
F.~Werner, L.~Tarruell, and Y.~Castin, ``Number of closed-channel molecules in the BEC-BCS crossover''  \href{https://doi.org/10.1140/epjb/e2009-00040-8}{ \emph{Eur. Phys. J. B}, {\bfseries ~68},  401--415, 2009}.
\BIBentrySTDinterwordspacing

\bibitem{Liu2021}
\BIBentryALTinterwordspacing
X.-P. Liu, X.-C. Yao, H.-Z. Chen, X.-Q. Wang, Y.-X. Wang, Y.-A. Chen, Q.~Chen, K.~Levin, and J.-W. Pan, ``{Observation of the density dependence of the closed-channel fraction of a 6Li superfluid}''  \href{https://doi.org/10.1093/nsr/nwab226}{ \emph{National Science Review}, {\bfseries ~9},  nwab226, 2021}.
\BIBentrySTDinterwordspacing

\bibitem{Li2022}
\BIBentryALTinterwordspacing
Y.~Li, F.-K. Guo, J.-Y. Pang, and J.-J. Wu, ``Generalization of Weinberg's compositeness relations''  \href{https://link.aps.org/doi/10.1103/PhysRevD.105.L071502}{ \emph{Phys. Rev. D}, {\bfseries  105},  L071502, 2022}.
\BIBentrySTDinterwordspacing

\bibitem{Goral2004}
\BIBentryALTinterwordspacing
K.~G\'{o}ral, T.~K\"{o}hler, S.~A. Gardiner, E.~Tiesinga, and P.~S. Julienne, ``Adiabatic association of ultracold molecules via magnetic-field tunable interactions''  \href{https://dx.doi.org/10.1088/0953-4075/37/17/006}{ \emph{J. Phys. B}, {\bfseries ~37},  3457, 2004}.
\BIBentrySTDinterwordspacing

\bibitem{Julienne2006}
\BIBentryALTinterwordspacing
P.~S. Julienne and B.~Gao, ``{Simple Theoretical Models for Resonant Cold Atom Interactions}''  \href{https://doi.org/10.1063/1.2400656}{ \emph{AIP Conference Proceedings}, {\bfseries  869},  261--268, 2006}.
\BIBentrySTDinterwordspacing

\bibitem{Nygaard2006}
\BIBentryALTinterwordspacing
N.~Nygaard, B.~I. Schneider, and P.~S. Julienne, ``Two-channel $R$-matrix analysis of magnetic-field-induced Feshbach resonances''  \href{https://link.aps.org/doi/10.1103/PhysRevA.73.042705}{ \emph{Phys. Rev. A}, {\bfseries ~73},  042705, 2006}.
\BIBentrySTDinterwordspacing

\bibitem{Kraats2023}
\BIBentryALTinterwordspacing
J.~van~de Kraats, D.~J.~M. Ahmed-Braun, J.-L. Li, and S.~J. J. M.~F. Kokkelmans, ``Efimovian three-body potential from broad to narrow Feshbach resonances''  \href{https://link.aps.org/doi/10.1103/PhysRevA.107.023301}{ \emph{Phys. Rev. A}, {\bfseries  107},  023301, 2023}.
\BIBentrySTDinterwordspacing

\bibitem{Naidon2017}
\BIBentryALTinterwordspacing
P.~Naidon and S.~Endo, ``Efimov physics: a review''  \href{https://dx.doi.org/10.1088/1361-6633/aa50e8}{ \emph{Rep. Prog. Phys.}, {\bfseries ~80},  056001, 2017}.
\BIBentrySTDinterwordspacing

\bibitem{Xie2020}
\BIBentryALTinterwordspacing
X.~Xie, M.~J. Van~de Graaff, R.~Chapurin, M.~D. Frye, J.~M. Hutson, J.~P. D'Incao, P.~S. Julienne, J.~Ye, and E.~A. Cornell, ``Observation of Efimov Universality across a Nonuniversal Feshbach Resonance in $^{39}\mathrm{K}$''  \href{https://link.aps.org/doi/10.1103/PhysRevLett.125.243401}{ \emph{Phys. Rev. Lett.}, {\bfseries  125},  243401, 2020}.
\BIBentrySTDinterwordspacing

\bibitem{Secker2021}
\BIBentryALTinterwordspacing
T.~Secker, D.~J.~M. Ahmed-Braun, P.~M.~A. Mestrom, and S.~J. J. M.~F. Kokkelmans, ``Multichannel effects in the Efimov regime from broad to narrow Feshbach resonances''  \href{https://link.aps.org/doi/10.1103/PhysRevA.103.052805}{ \emph{Phys. Rev. A}, {\bfseries  103},  052805, 2021}.
\BIBentrySTDinterwordspacing

\bibitem{Yudkin2023}
\BIBentryALTinterwordspacing
Y.~Yudkin, R.~Elbaz, J.~P. D'Incao, P.~S. Julienne, and L.~Khaykovich, ``The reshape of three-body interactions: Observation of the survival of an Efimov state in the atom-dimer continuum''  \href{https://doi.org/10.1038/s41467-024-46353-1}{ \emph{Nat Commun}, {\bfseries ~15},  2127, 2024}.
\BIBentrySTDinterwordspacing

\bibitem{Kraats2023a}
\BIBentryALTinterwordspacing
J.~van~de Kraats, D.~J.~M. Ahmed-Braun, J.-L. Li, and S.~J. J. M.~F. Kokkelmans, ``Emergent Inflation of the Efimov Spectrum under Three-Body Spin-Exchange Interactions''  \href{https://link.aps.org/doi/10.1103/PhysRevLett.132.133402}{ \emph{Phys. Rev. Lett.}, {\bfseries  132},  133402, 2024}.
\BIBentrySTDinterwordspacing

\end{thebibliography}

\end{document}